\documentclass[twocolumn,floatfix,nofootinbib,amsmath,amssymb,
 aps,prd]{revtex4-1}

\usepackage{graphicx}
\usepackage{dcolumn}
\usepackage{bm}
\usepackage{color}

\def\mnras{MNRAS}
\def\jcap{JCAP}

\begin{document}


\title{Strong-field gravitational-wave emission in Schwarzschild and Kerr geometries: some general considerations}

\author{J.~F.~Rodr\'iguez, J.~A.~Rueda, R.~Ruffini}
\email{jose.rodriguez@icranet.org;jorge.rueda@icra.it;ruffini@icra.it}
\affiliation{%
ICRANet, Piazza della Repubblica 10, I--65122 Pescara, Italy\\
Dipartimento di Fisica and ICRA, Sapienza Universit\`a di Roma, P.le Aldo Moro 5, I--00185 Rome, Italy
}

\date{\today}

\begin{abstract}
We show how the concurrent implementation of the exact solutions of the Einstein equations, of the equations of motion of the test particles, and of the relativistic estimate of the emission of gravitational waves from test particles, can establish a priori constraints on the possible phenomena occurring in Nature. Two examples of test particles starting at infinite distance or from finite distance in a circular orbit around a Kerr black hole are considered: the first leads to a well defined gravitational wave burst the second to a smooth merging into the black hole. This analysis is necessary for the study of the waveforms in merging binary systems.
\end{abstract}

\maketitle


\section{Introduction}\label{sec:1}

The field of relativistic astrophysics has been flourished taking advantage on the developments in the physics of neutron stars and of black holes thanks to the discovery of quasars \cite{1963Natur.197.1040S}, of pulsars \cite{1968Natur.217..709H}, of binary X-ray sources \cite{1972ApJ...178..281G,1973ApJ...180L..15L,1978pans.proc.....G}, of gravitational waves from binary neutron stars \cite{1975ApJ...195L..51H,2005ASPC..328...25W}, and of gamma-ray bursts \cite{1975ASSL...48.....G}.

{From the above experience one can recognize two crucial aspects of an experimental discovery: 1) the reproducibility of the experimental data and 2) the consistency of the data with the physical laws and the theoretical treatment. In view of the recent boosted interest on gravitational waves due to claimed ``observation of gravitational waves from a binary black hole merger'' by the LIGO-Virgo Collaboration \cite{2016PhRvL.116f1102A,2016PhRvL.116x1103A,2017PhRvL.118v1101A}, we focus here on the second aspect and consider some theoretical issues concerning the gravitational-wave emission of a merging black-hole binary that in our opinion deserve attention.}

We limit ourselves to the case of perturbations in the strong-field limit. We are going to neglect all post-Newtonian approaches to this problem (see Sec.~\ref{sec:4} for a brief discussion on this issue). Two idealized processes have been introduced in the strong-field limit to describe the motion of a test particle plunging, following a geodesic, into the field of an already formed black hole. The first case is the one of a particle initially at rest or with a finite kinetic energy plunging from infinite distance. The second is the one of a particle initially in circular orbit and finally merging into a black hole. Both the Schwarzschild metric and the Kerr metric for a black hole have been considered. This transition from the Schwarzschild to the Kerr case has been addressed for almost 50 years. Broadly speaking, all particles starting from infinite distance give rise to ``\emph{a precursor, a main burst and a ringing tail}''. A finite energy emission occurs, $\Delta E = \eta \mu (\mu/M)$, where $\mu$ and $M$ are the particle and black-hole mass, with $\eta$ ranging from 0.01 to almost 1 in the fully general relativistic treatment, as a function of the nature of the black hole and of the initial energy and angular momentum of the particle. We use throughout geometric units with $c=G=1$.

Circular orbits are equally interesting. It has been relevant the introduction of the ``effective potential technique'' for the understanding of the general properties of these geodesics, ranging from infinite distance all the way to the last circular orbit (hereafter LCO), as well as the approach to the black-hole horizon. Since unstable orbits are not physically relevant, i.e. they are not realizable in nature, we refer to as circular orbits the only ones that can exist, namely the stable ones. The LCO follows then the traditional definition \cite{1971ESRSP..52...45R}, quoted in \cite{1975ctf..book.....L} (problem 2 of \$104), as the inflection point of the effective potential (see Sec.~\ref{sec:3}) and it is the circular orbit closest to the black-hole horizon.

As we are going to show in this article, the gravitational-wave emission leads to a ``helicoidal drifting sequence'' (hereafter HDS) of orbits around the Kerr black hole, giving rise to a smooth transition to the plunging phase into the black hole without a final burst. {We have recently used the treatment presented in this work to perform a comparison of numerical relativity binary black-hole merger waveforms of the SXS catalog \cite{SXS:catalog} with HDS waveforms up to the LCO \cite{2018JCAP...02..030R}. We have found, until the LCO, an unexpected and yet theoretically unexplained agreement, between both waveforms in the comparable-mass regime for spinless, aligned as well as anti-aligned merging black-hole binaries, for equal and unequal values of the binary mass-ratio.}

{Concerning the plunge phase, we show here that our results differ with the ones in the literature, e.g. Ref.~\cite{2000PhRvD..62l4022O}, that show a larger amount of energy radiated in gravitational waves during the plunge into the black hole. A comparable energetic plunge leading to a burst of radiation in the black hole binary merger appears also in the numerical-relativity waveforms of the SXS catalog \cite{SXS:catalog}. Such a feature is also found in the binary coalescence process modeled via the effective one-body (EOB) formalism \cite{1999PhRvD..59h4006B} which adopts a treatment as the one in \cite{2000PhRvD..62l4022O} for the plunge phase (see Ref.~\cite{2000PhRvD..62f4015B} for details). In view of the large use of these waveforms by the LIGO-Virgo Collaboration for the binary parameter estimation, see e.g. the case of the GW150914 event \cite{2016PhRvL.116f1102A}, these results are far from being just an academic exercise. In this line it is also important to recall a most important recent result of an independent analysis of GW150914 that shows the incompatibility of the LIGO-Virgo data with the presence of such a burst, in clear contrast with the currently used waveform templates (see, e.g., Figs.~5 and 10 in Ref.~\cite{2018JCAP...02..013L} and also Refs.~\cite{2016JCAP...08..029N,2016JCAP...10..014L,2017JCAP...08..013C}, for further details).}

The article is organized as follows. We recall in Sec.~\ref{sec:2} the main results on the gravitational-wave emission of particles falling into a black hole from infinite distance. In Sec.~\ref{sec:3} we recall the results on the case of circular orbits without taking into account radiation reaction and show the gravitational-wave energy and angular momentum flux at infinity, following the Sasaki-Nakamura method. Section~\ref{sec:4} is devoted to the formulation of the equations of motion of the HDS of the test particle, taking into account the radiation reaction. In Sec.~\ref{sec:5} we discuss the numerical results for specific examples of the evolution up to the passage of the particle at the location of the LCO. In Sec.~\ref{sec:6} we discuss the plunging of the particle into the black hole. We explicitly show the incongruence of some results in the literature (see e.g. Ref.~\cite{2000PhRvD..62l4022O}). Finally, we present in Sec.~\ref{sec:6} our conclusions.

\section{Infall of a test particle into a black hole starting at infinite distance}\label{sec:2}

That gravitational radiation can be emitted by a test particle falling radially into a Schwarzschild black hole was shown in a simple computation assuming that the particle follows a geodesic and describing the radiation in flat spacetime in a linearized theory \cite{RuffiniWheeler71} (see also Ref.~\cite{1974bhgw.book.....R}). Both an estimate of the energy emitted, $\Delta E = 0.0025 \mu (\mu/M)$, and of the gravitational-wave spectrum were there presented. In \cite{1970PhRvL..24..737Z,1970PhRvD...2.2141Z} it was introduced a mathematically more advanced treatment using the decomposition of the perturbation into the tensorial spherical harmonics \cite{1957PhRv..108.1063R} reaching a second-order linear equations: the Zerilli equation.

The numerical integration of this equation by the Green-function technique in \cite{1971PhRvL..27.1466D} led to an improvement by a factor of 6 the previous estimate in \cite{RuffiniWheeler71} and by a factor of 4 the one in \cite{1970PhRvL..24..737Z,1970PhRvD...2.2141Z}. It also allowed the determination of the multipole components of the spectrum. This, in turn, allowed soon after the derivation of the associated gravitational-wave pulse composed with three components, ``\emph{a precursor, a main burst and a ringing tail}'' (see Fig.~1 in Ref.~\cite{1972PhRvD...5.2932D}). Still using the Regge-Wheeler-Zerilli approach, in \cite{1973PhRvD...7..972R} it was studied the gravitational-wave radiation by a particle projected from infinity with finite kinetic energy into a Schwarzschild black hole. It was found a difference in the spectrum in the low-frequency region and a larger amount of radiation emitted by increasing the Lorentz gamma-factor of the injected particle. 

To these works a vast activity followed, on one side the extension of the Ruffini-Wheeler formulation of a particle thrust into a Schwarzschild black hole \cite{1981PThPh..66.1627R}, and showing a corresponding increase of the burst structure. On other side, a ``long march'' started following a pioneering work \cite{1973ApJ...185..635T}, which allowed to extend the above considerations to be extended to the case of particles, still starting at infinite distance, and now moving in the field of a Kerr black hole. A first step was initiated in \cite{1979ApJ...231..211D} by analyzing a particle plunging with finite angular momentum in the field of a Schwarzschild black hole. This work was based on techniques introduced by Teukolsky and further elaborated \cite{1976RSPSA.350..165C}. In \cite{1979ApJ...231..211D} it was shown that the energy emitted may be enhanced by a factor of 50 with respect to the radial infall. They confirmed that the gravitational-wave burst was still composed of the above three components. The further development needed a fundamental progress in the perturbation technique introduced in \cite{1982PhLA...89...68S,1982PThPh..67.1788S}. By introducing a change of variable in the Teukolsky radial equation in order to have a short-range potential, they produced an entire family of new results. Among these, the gravitational emission of a test particle infalling along the rotational axis of an almost extreme Kerr black hole with spin parameter $a/M = 0.99$ ( $0\leq a \equiv J_{\rm BH}/M \leq 1$, where $J_{\rm BH}$ and $M$ are, respectively, the black-hole angular momentum and mass), obtaining a new multipole distribution of the radiation closely following the above three components, and the total energy radiated $0.0170 \mu (\mu/M)$, i.e. a factor 1.65 larger than the energy radiated found in \cite{1971PhRvL..27.1466D}. A second important step was made in \cite{1984PThPh..71...79K} by studying the gravitational waves from a particle with non-zero orbital angular momentum plunging on the equatorial plane of a Kerr black hole, and proving that the energy radiated, for a Kerr black hole of $a/M\sim 1$, could reach the limiting case $\Delta E \sim \mu (\mu/M)$. We shall come back to this point in Sec.~\ref{sec:6} where the transition of a particle from the inspiral, quasi-circular phase, to the final plunge into the black hole, is analyzed. We shall estimate the amount of energy radiated to infinity in such a transition and compare with some existent treatment in the literature, e.g. the one in \cite{2000PhRvD..62l4022O}.

In conclusion, all particles plunging or thrusting from infinite distance into both a Schwarzschild and a Kerr black hole, present a characteristic burst composed of ``\emph{a precursor, a main burst and a ringing tail}''.

\section{The case of circular orbits around the black hole}\label{sec:3}

The first analysis for a relativistic treatment for the radiation in circular orbit around a Schwarzschild black hole was motivated by the declaration in \cite{1969PhRvL..22.1320W} of the discovery of gravitational waves and, by the contention of their explanation in terms of synchrotron gravitational radiation \cite{1972PhRvL..28..994M} (see also Ref.~\cite{1972PhRvL..28..998M}). For this reason, the multipole modes of gravitational radiation by a particle moving in a circular orbit was studied \cite{1972PhRvL..28.1352D} and it was found that the enhancement of high-multipoles, indicated in Ref.~\cite{1972PhRvL..28..994M}, does not exist and consequently that no synchrotron gravitational radiation can occur.

Paradoxically, this treatment of circular orbits in the Schwarzschild metric was reproduced, without due reference in \cite{1993PhRvD..47.1511C}, and also in \cite{2009PhRvD..79f4004D} who used as the ``exact result'' over which construct their model connecting the post-Newtonian treatment to the strong-field regime. The extension of this treatment for eccentric orbits can be found in Ref.~\cite{1993PThPh..90...65T}.

The problem of circular orbits in the field of a Kerr black hole was formulated in \cite{1978ApJ...225..687D} formulated. The fundamental works \cite{1993PhRvD..48..663S,1993PThPh..90..595S} treat the same physical problem but within the Sasaki-Nakamura treatment.

We compare and contrast here the results in the Schwarzschild metric \cite{1972PhRvL..28.1352D} with the ones developed in the Kerr metric. It is appropriate to clarify that the considerations in this section do address the gravitational energy emitted by a circular orbit, initially neglecting all effects of radiation reaction. Therefore, we consider idealized circular orbits on the equatorial plane, at constant radii, including the LCO, originally introduced in \cite{1971ESRSP..52...45R} in the analysis of the radial effective potential. We recall its expression \cite{1971ESRSP..52...45R} (reproduced in \cite{1974bhgw.book.....R}):
\begin{equation}\label{eq:Veff}
V_{\rm eff} =1-\frac{2 M}{r}+\frac{l^2-a^2 (\epsilon^2-1)}{r^2}-\frac{2 M (l-a \epsilon)^2}{r^3},
\end{equation}
which leads to a radial equation of motion
\begin{equation}
\epsilon^2 = \left(\frac{dr}{d\tau}\right)^2 + V_{\rm eff},
\end{equation}
where $\epsilon \equiv E/\mu$ and $l \equiv L/\mu$ are the particle's energy and angular momentum per unit mass, and $\tau$ is the proper time.

We are here interested in corotating circular orbits (obtained from the conditions $dr/d\tau= 0$ and $\partial V_{\rm eff}/\partial r = 0$), which have energy and orbital angular momentum given by \cite{1971ESRSP..52...45R,1974bhgw.book.....R}
\begin{align}
\epsilon &= \frac{E}{\mu} = \frac{r^2 - 2 M r + a M^{1/2} r^{1/2}}{r (r^2 - 3 M r + 2 a M^{1/2} r^{1/2})^{1/2}}, \label{eq:Eisco}\\
\frac{l}{M} &= \frac{L}{\mu M} = \frac{r^2 - 2 a M^{1/2} r^{1/2} + a^2}{r^{3/4} (r^{3/2} - 3 M r^{1/2} + 2 a M^{1/2})^{1/2}}.\label{eq:Lisco}
\end{align}

The LCO is given by the inflection point of the effective potential, i.e. the radius for which $\partial^2 V_{\rm eff}/\partial r^2 = 0$. In the case of $a/M=0$ (Schwarzschild metric), it is located at $r_{\rm LCO} = 6 M$ and $E_{\rm LCO}/\mu = 2\sqrt{2}/3$ and $L_{\rm LCO}/(\mu M) = 2\sqrt{3}$. In the case of an extreme Kerr black hole, $a/M = 1$, the LCO is located very close (but not coincident) to the black-hole horizon, i.e. $r_{\rm LCO}\to r_{+}$ where $r_{+} = M$, and $E_{\rm LCO}/\mu = \sqrt{3}/3$ and $L_{\rm LCO}/(\mu M) = 2\sqrt{3}/3$. Namely, for an extreme black hole there exist circular orbits up to very close to the black-hole horizon (see e.g. Ref.~\cite{1972ApJ...178..347B} for additional details).

\begin{figure*}
\centering
\includegraphics[width=\hsize,clip]{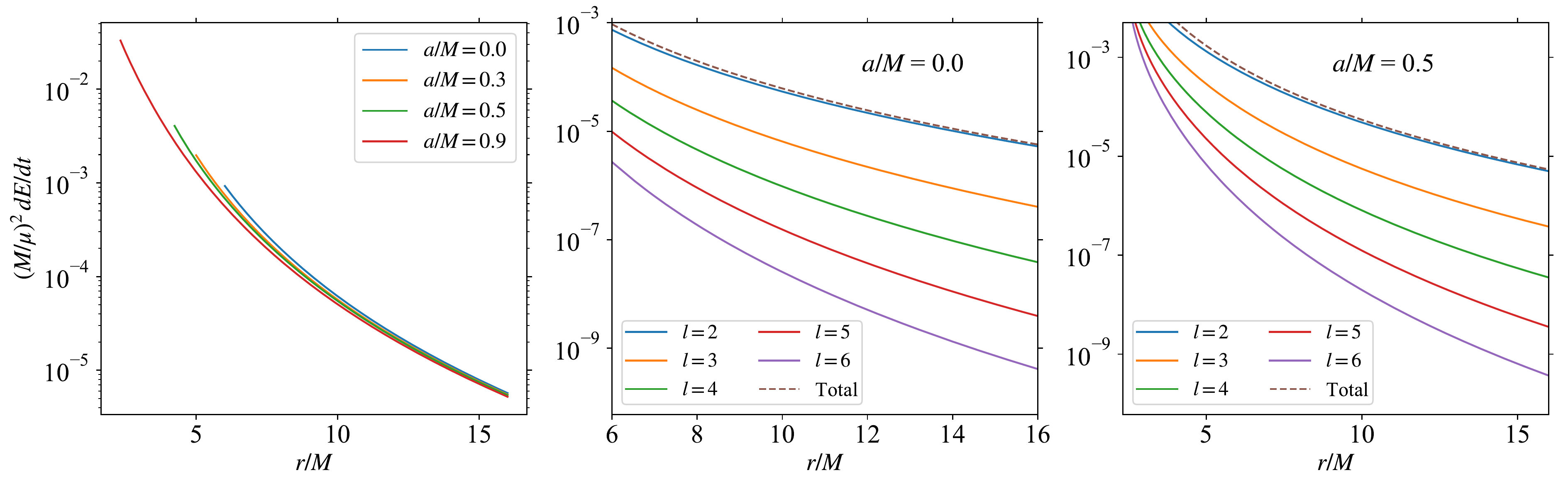}
\caption{Left: comparison of the total gravitational-wave energy flux at infinity, $\dot{E} \equiv dE/dt$, emitted by the test particle in circular orbits around the black hole for selected values of the black-hole dimensionless spin, as a function of the dimensionless radial position, $r/M$. Center and Right: contribution of the gravitational-wave modes $l \geq 2$ ($m = -l$) to the total gravitational-wave energy flux at infinity in the case of a test particle in circular orbits of radius $r$ around a Schwarzschild black hole (center panel) and around a Kerr black hole with $a/M = 0.5$ (right panel). It can be seen that the $(2,2)$ highly dominates the gravitational-wave emission at any radius down to the marginally unstable orbit.
}\label{fig:dEdt}
\end{figure*}

Now we turn to the calculation of the energy and momentum fluxes, it can be done by using standard metric perturbation theory, e.g. the pioneering work of Regge-Wheeler for perturbations in the Schwarzschild spacetime. However, there is an alternative approach developed by Teukolsky, involving curvature perturbations instead of metric perturbations  in which the scalar, vector and tensor perturbations are governed by a single master equation. The master equation can be separated and the solution is an expansion in  fourier and spheroidal harmonic modes \cite{1972PhRvL..29.1114T, 1973ApJ...185..635T, 1974ApJ...193..443T}. Unfortunately, the Teukolsky radial equation has a long-range potential and its numerical integration with boundary conditions is difficult. In \cite{1982PThPh..67.1788S,1982PhLA...89...68S} it was found a change of variables that introduces a short-range and well-behaved potential $U(r)$. The Sasaki-Nakamura equation is:
\begin{equation}\label{eqn:sasaki-nakamura}
	X''_{lm\omega} - F(r) X'_{lm\omega} - U(r) X_{lm\omega} = \mathcal{S}_{lm\omega}.
\end{equation}
Details on the functions $F$, $U$ and $S_{lm\omega}$ and their numerical solution can be found in \cite{1982PhLA...89...68S} (see, also \cite{2000PhRvD..61h4004H,nakano2016}). The task is accomplished by first solving numerically the Eq.~\eqref{eqn:sasaki-nakamura} with the suitable boundary conditions, and then inverting the transformation to find the original radial function. The solution of the Eq.~\eqref{eqn:sasaki-nakamura} is obtained by using the Green's function technique for boundary value problems. The two solutions of the homogeneous Sasaki-Nakamura equation are $X_{lm\omega}^H$, which satisfies the boundary condition of in-going radiation at the outer horizon, and $X_{lm\omega}^{\infty}$ which satisfies the boundary condition of out-going radiation at $\infty$.

In the case of circular orbits, the energy and angular momentum fluxes carried by the gravitational waves to infinity are given by 
\begin{align}
 \frac{dE}{dt} &= \sum_{l \geq 2}^{\infty}\sum_{m=-l}^{l} \frac{|Z_{lm\omega}^H|^2}{4\pi \omega^2_m},\label{eq:Edot}\\ 
\frac{d J}{dt} &= \sum_{l\geq 2}^{\infty}\sum_{m=-l}^{l} \frac{m |Z_{lm\omega}^H|^2}{4\pi \omega^3_m},\label{eq:Jdot}
\end{align}
where $\omega_m = m \Omega$, and $Z_{lm\omega}^H$ is a complex number which depends on the orbital frequency $\Omega$ and it is obtained from the solution $X_{lm\omega}^H$, see Ref.~\cite{2000PhRvD..61h4004H} for the explicit expression. To summarize, all that is needed to find the fluxes at infinity are the complex numbers $Z_{lm\omega}^H$. Without introducing further technical details which can be found in all the aforementioned references, we turn now to present the numerical results. Fig.~\ref{fig:dEdt} shows the total energy flux $dE/dt$ for selected values of the black-hole dimensionless spin. 

We show also in Fig.~\ref{fig:dEdt} the contribution of the gravitational-wave modes $l \geq 2$  to the total gravitational-wave energy flux at infinity. We explicitly show that enhancement of higher multipoles does not occur either in the case of the Kerr metric, where also the quadrupole contribution is largely predominant. No synchrotron gravitational radiation can occur either in the Schwarzschild or the Kerr case. The contribution of higher multipoles becomes relevant only for near-horizon orbits around nearly extremal black holes \cite{2016CQGra..33o5002G}.

\section{Helicoidal drifting sequence}\label{sec:4}

The Hamiltonian of the test particle of mass $\mu$ in the field of the Kerr black hole of mass $M$ is given by (see e.g. Ref.~\cite{1992AnPhy.215....1J}, and references therein)
\begin{equation}
H = - p_t = - N^i p_i +  N \sqrt{\mu^2+\gamma^{i j} p_i p_j},
\end{equation}
where $N = 1/\sqrt{-g^{00}}$, $N^i = - g^{t i}/g^{tt}$ and $\gamma^{i j} = g^{ij} + N^i N^j/N^2 = g^{i j}-g^{t i} g^{t j}/g^{tt}$. Here Latin index stands for the spatial Boyer-Lindquist coordinates $(r, \theta, \phi)$, $p_r$ and $p_\phi$ are, respectively, the radial and the angular momentum of the particle.

The Hamilton canonical equations \emph{for a test particle} on the equatorial plane $\theta=\pi/2$ under the action of radial and azimuthal dissipative effects can then be written as
%
\begin{align}
	\frac{dr}{dt} &=  \frac{\partial H}{\partial p_r},\label{eq:rdot}\\
	\frac{d\phi}{dt} &\equiv \Omega = \frac{\partial H}{\partial p_\phi},\label{eq:phidot}\\
	\frac{d p_r}{dt} &= -\frac{\partial H}{\partial r} + {\cal F}_r^{\rm nc},\label{eq:prdot}\\
	\frac{d p_\phi}{dt} &= {\cal F}_\phi^{\rm nc},\label{eq:pphidot}
\end{align}
{We consider here only the non-conservative part of the radiation-reaction force (see e.g. Ref.~\cite{2011LRR....14....7P} for a review on the subject), so} we adopt for the radial and azimuthal non-conservative radiation-reaction forces:
\begin{align}
	{\cal F}_r^{\rm nc} &= 0,\label{eq:Fr}\\
	{\cal F}_\phi^{\rm nc} &= -\frac{d J}{dt}.\label{eq:Ldot}
\end{align}
where $J$ is the angular momentum carried out to infinity by the gravitational waves given by Eq.~(\ref{eq:Jdot}). The assumption (\ref{eq:Fr}) is supported from previous results that show that the linear momentum carried out by the waves to infinity satisfies \cite{1984MNRAS.211..933F}
\begin{equation}\label{eq:Pdot}
\frac{d p^{\rm GW}_r}{dt} \ll \left|\frac{\partial H}{\partial r}\right|.
\end{equation}
We recall that for purely quadrupolar waves in strict circular orbit the following equality is satisfied:
\begin{equation}\label{eq:Jdot2}
	\frac{d J}{dt} = \frac{1}{\Omega}\frac{dE}{dt}.
\end{equation}
However, since we have the presence of the radial drift and a small contribution of higher order multipoles, the above equality must not be strictly satisfied (see Eqs.~\ref{eq:Edot}--\ref{eq:Jdot}). We have checked from our numerical computations that $|1-\Omega (dJ/dt)/(dE/dt)|\sim 10^{-6}$ during the evolution. The smallness of this value imply that the motion is indeed quasi-circular, but it is sufficiently large (with respect to the numerical precision of our calculations) to conclude that the equality (\ref{eq:Jdot2}) is not verified in the HDS evolution, as expected.

It has been traditional to treat the HDS evolution using the so-called adiabatic approximation by assuming that the particle moves from a circular orbit to the next (see, e.g., Ref.~\cite{2000PhRvD..62l4021F,2001PhRvD..64f4004H}),due to the energy and angular momentum radiation. For example, since the energy can be expressed in terms of $r$, a change in the energy induces a change in $r$.
Thus, it can be found a dynamical equation for $r$ by differentiating \eqref{eq:Eisco} and equating it to the gravitational energy flux. 
The evolution is found by integrating this equation instead of solving the equations of motion (\ref{eq:rdot})--(\ref{eq:pphidot}).
This approximation, although sufficient to estimate some general properties of the quasi-circular evolution, lacks an appropriate inclusion of the non-zero radial motion of the particle in the equations of motion. As we shall show below, this radial drift becomes essential to have the correct initial conditions for the computation of the evolution of the particle beyond the HDS in the final plunge to the black hole. When the radial momentum is properly included, the LCO location does not represent any longer a point where the equations of motion break down. Indeed, the conditions of ``adiabaticity'' are kept up to such distances (see Fig.~\ref{fig:adiabatic} and related discussion in Sec.~\ref{sec:5}).

It is appropriate to recall at this point some other treatments in the literature on the problem of a binary system with a test particle component evolving under the gravitational-wave emission.  
In \cite{1999PhRvD..59h4006B} it was proposed the EOB approximation to overcome the known problem of the non-convergence of the higher-order post-Newtonian successive approximations. The EOB treatment ``maps'' the post-Newtonian binary into a ``Schwarzschild deformed metric'' which depends on the mass-ratio. In the extreme mass-ratio limit $\mu/M \to 0$, the metric becomes the Schwarzschild one and the dynamics becomes the one of a test particle around a Schwarzschild black hole. However, the treatment of the waveform and/or the gravitational-wave fluxes stands on a post-Newtonian basis. The Kerr black-hole metric in the EOB treatment has been used to treat spinning merging components \cite{2001PhRvD..64l4013D}. However, also that EOB treatment is based on a post-Newtonian treatment for the waveforms and/or fluxes. Thus although ``calibrated'' to fit strong-field results, the EOB approach remains conditioned by the non-convergence of the post-Newtonian formalism.

{The set of equations of motion (\ref{eq:rdot}--\ref{eq:pphidot}) was first used in \cite{2011PhRvD..84d4014H} also using the radiation-reaction term obtained from the numerical solution of the Teukolsky equation for circular orbits. However, they analyze the case of intermediate mass-ratios with the conservative dynamics given by the aforementioned EOB treatment.}

{There has been also introduced a different method to compute the evolution of accelerated orbits that is based on linking one reference geodesic to the next by describing changes in the orbital ``constants'' of motion \cite{2008PhRvD..77d4013P}. Such reference geodesics, called ``osculating orbits'', with a planar force have been applied to the inspiral of a particle in the Schwarzschild background in \cite{2008PhRvD..77d4013P} and in the Kerr background in \cite{2011PhRvD..83d4037G}. Since we do not include here the radial velocity in the calculation of the gravitational-wave flux driving the inspiral, namely we use the one of an exact circular geodesic orbit, our scheme and the osculating orbits one become equivalent under these assumptions. However, we here use the fully relativistic gravitational-wave flux instead of a post-Newtonian as in \cite{2008PhRvD..77d4013P}.}

{It has been also often applied the ``adiabatic'' approximation (see e.g.~Ref.~\cite{2000PhRvD..62l4022O}) in which the inspiral motion of the test particle is modeled by connecting one real (i.e.~not osculating) circular geodesic to the next by using flux-balance, i.e. $dr/dt = (dr/dE) (dE/dt)$, with $E$ the energy of the particle in the circular orbit around the black hole (see Eq.~\ref{eq:Eisco}). However, such an adiabatic approximation breaks down near the location of the LCO and thus needs the introduction of a separate treatment of the transition from the inspiral to the plunge phase (see Ref.~\cite{2000PhRvD..62l4021F,2001PhRvD..64f4004H} and Sec.~\ref{sec:5} for details).}

More recently, the problem of a test particle inspiraling into a Kerr black hole was treated \cite{2014PhRvD..90h4025T} in a similar way as we do here but with a drawback. The gravitational-wave flux was there adopted after the LCO as given by the Teukolsky equation integrated for the unstable circular orbits up to the light ring.

On the other hand, full numerical-relativity simulations are available only for relatively large mass-ratios $\gtrsim 1/10$ as it can be seen from the SXS catalog \cite{SXS:catalog} of binary black-hole merger simulations performed with the Spectral Einstein Code (SpEC). Therefore, it is not currently possible to perform a one-to-one comparison between numerical-relativity simulations and the test-particle treatment. {However, we have recently performed a comparison of the SXS waveforms with the ones obtained for the HDS dynamics described in the present article, in the comparable-mass regime \cite{2018JCAP...02..030R}. Unexpectedly, we have found a great agreement of both waveforms for spinless, as well as for aligned and anti-aligned merging black hole binaries, for equal and unequal values of the binary mass-ratio.}

\section{Numerical results}\label{sec:5}

We integrate Eqs.~(\ref{eq:rdot})--(\ref{eq:pphidot}) numerically with given initial conditions appropriate for quasi-circular orbits on the equatorial plane. At the initial time $t_0=0$ we set the system at an initial distance $r_0$, the initial phase is $\phi(t_0,r_0)=0$. The angular momentum is  $p_\phi(t_0,r_0) = L_0$, where $L_0 \equiv L(r_0)$ is given by Eq.~(\ref{eq:Lisco}). The initial condition for the radial momentum can be obtained from the equations of motion as follows. The radiative force induces a radial velocity
{
\begin{equation}\label{eq:Pr0}
\left.\left(\frac{dp_\phi}{dt}\right)\right|_{t_0,r_0} = \left.\left(\frac{dp_\phi}{dr}\right)\left(\frac{dr}{dt}\right)\right|_{t_0,r_0} = \left.\mathcal{F}^{\rm nc}_{\phi}\right|_{r_0}.
\end{equation}
}

The non-zero radial velocity is related to $p_{r}$ via Eq.~\eqref{eq:rdot} which when introduced into the above expression leads to a non-linear algebraic equation for $p_{r}|_{r_0}$. Such equation can be solved numerically given all the above conditions and, at leading order, it can be solved analytically giving rise to:
\begin{equation}
p_{r}\bigr\rvert_{r_0} = \sqrt{\biggl(\mu^2 + \frac{r_0^2 L^2_0}{\Lambda_0}\biggr)\frac{r_0^2\Delta_0}{\Lambda_0}}\frac{\Lambda_0\mathcal{F}_{\phi}^{\textrm{nc}}\bigr\rvert_{r_0}}{dL/dr_0(r_0^2 + a^2)^2},
\end{equation}
where $\Delta_0 = r_0^2 - 2 M r_0 + a^2$ and $\Lambda_0 = (r_0^2 + a^2)^2 - a^2\Delta_0$. This equation gives the initial condition for $p_r$ with very high accuracy and can be safely used providing the initial radius is sufficiently far from the radius of the LCO. For instance, for initial position $r_0 = 15.9 M$, Kerr black hole spin parameter $a/M=0.9$, mass-ratio $\mu/M = 1/100$, it gives the initial $p_r$ accurate within nine digits.

{This initial radial velocity condition, if given at a large enough value of $r_0$, reduces to one of the ``adiabatic'' approximation. First, from Eqs.~(\ref{eq:Pr0}) and (\ref{eq:Ldot}), we have that $dr/dt = - (dJ/dt)/(dp_\phi/dr)$. As we have mentioned, the property of a strict circular orbit, namely Eq.~(\ref{eq:Jdot2}), is satisfied in our system by one part in a million, therefore the condition $dr/dt = -  (1/\Omega)(dE/dt)/(dp_\phi/dr)$ is approximately satisfied with the same accuracy. Now, by replacing $\Omega$ via Eq.~(\ref{eq:phidot}), we finally obtain $dr/dt = (dE/dt)/(dE/dr)$, which is the flux-balance condition of the adiabatic approximation, adopted e.g. in \cite{2000PhRvD..62l4022O}.}

\begin{figure}
\centering
\includegraphics[width=\hsize,clip]{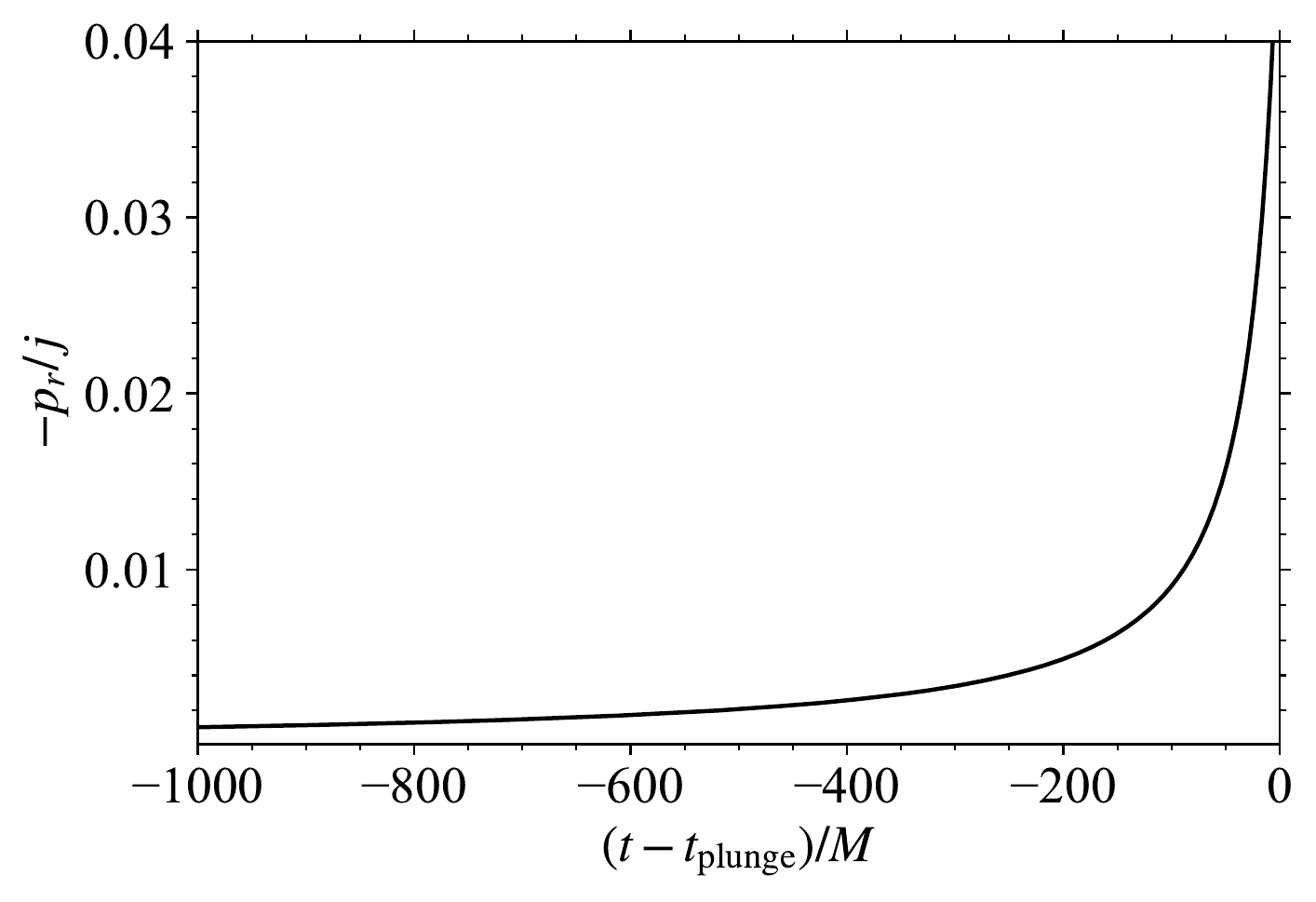}
\caption{$p_r/j$: ratio of the radial momentum to the angular momentum per unit mass of a test particle in the HDS around a Kerr black hole with $a/M = 0.9$. The mass-ratio is $\mu/M = 1/100$. The plot shows the ratio from $r~\approx 4.53 M$ up to the location of the LCO at $r\approx 2.32 M$. The time $t_{\rm plunge}$ is here defined as the time of the passage of the particle at the location of the LCO.}\label{fig:PrPphi}
\end{figure}

Figure~\ref{fig:PrPphi} shows the ratio of the radial momentum, $p_r$, to the angular momentum per unit mass, $j\equiv p_\phi/\mu$, during the HDS obtained for the afore-mentioned initial conditions. It can be seen how the importance of the radial momentum increases for shorter and shorter distances and is specially non-negligible near the location of the LCO. In this example, $p_r$ becomes $\sim 4\%$ of the angular momentum per unit mass.

\begin{figure}
\centering
\includegraphics[width=\hsize,clip]{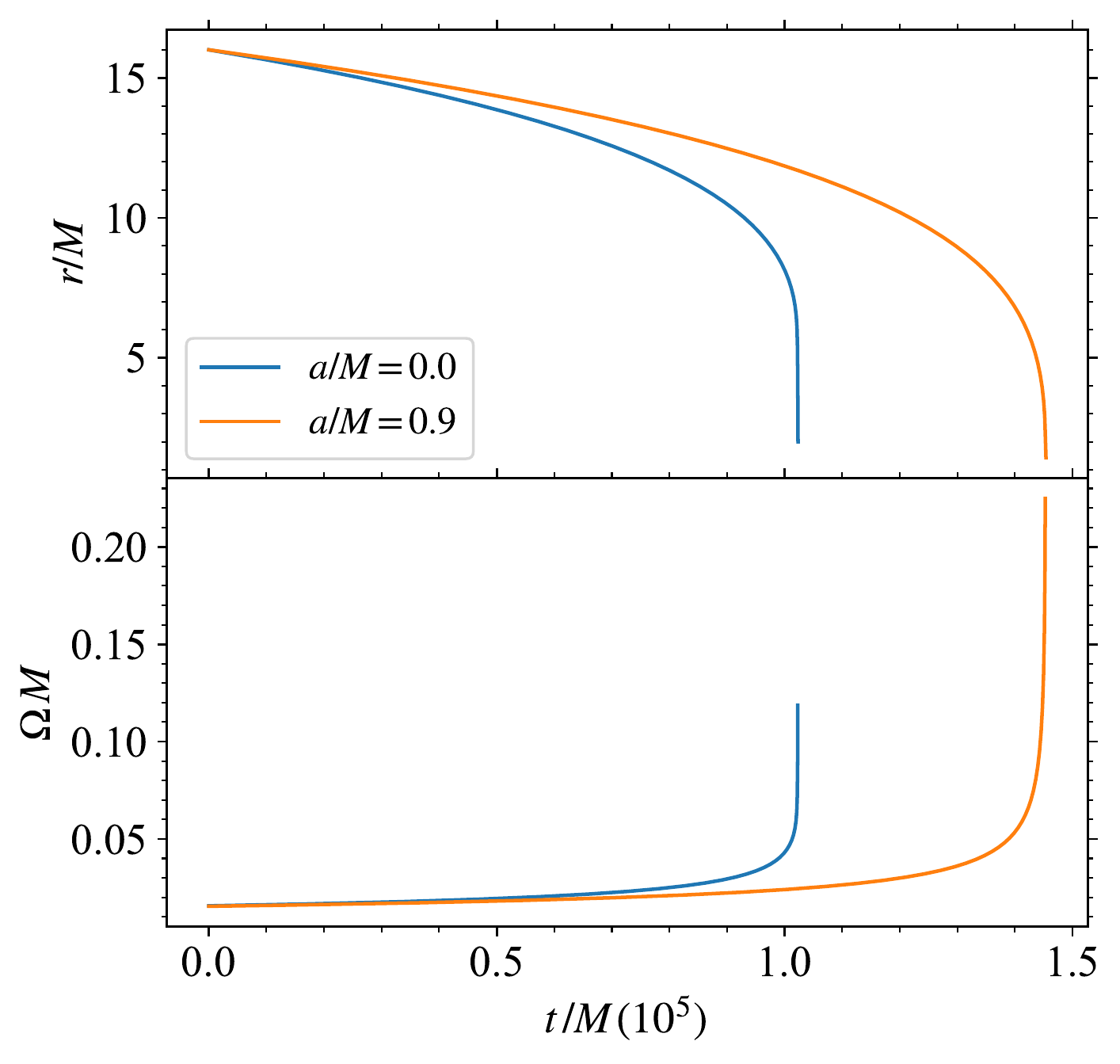}
\caption{The upper and lower panels show, respectively, the dimensionless radial position, $r/M$, and dimensionless orbital angular velocity, $\Omega M$, as a function of dimensionless time, $t/M$. The initial position at time $t/M = 0$ is $r/M = 15.9$ and the mass-ratio is $\mu/M = 1/100$. The case of a Schwarzschild black hole is shown by the blue-solid curves while the case of a Kerr black hole with $a/M = 0.9$ is shown by the red-dashed curves.
}\label{fig:romega}
\end{figure}

\begin{figure}
\centering
\includegraphics[width=\hsize,clip]{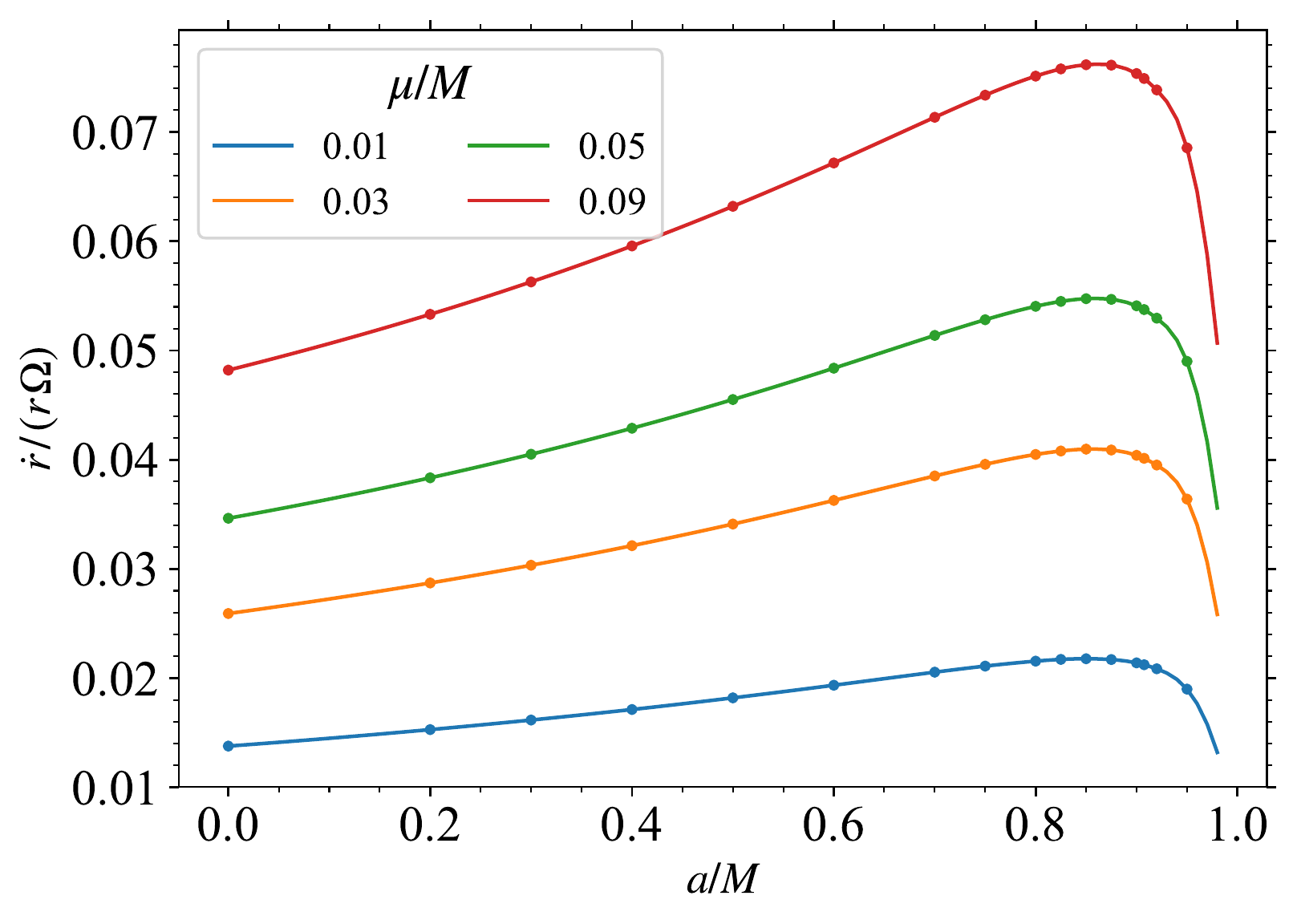}
\caption{Orbital to radiation-reaction timescale ratio $T_{\rm orb}/T_{\rm rad} = |\dot{r}|/(r \Omega)$ evaluated at the location of the LCO, for selected mass-ratios and selected values of the black-hole spin parameter.}\label{fig:adiabatic}
\end{figure}

Fig.~\ref{fig:romega} shows the dimensionless radial position, $r/M$, and dimensionless orbital angular velocity, $\Omega M$, as a function of dimensionless time, $t/M$. This plot shows the above quantities up to the LCO. The evolution in the final plunge into the black hole is discussed in the next section. The sharp decrease (increase) of $r$ ($\Omega$) with time near the location of the LCO suggests that the ``adiabaticity'' of the system might be loose at such distance. The system can be considered to evolve adiabatically if the orbital to radiation-reaction timescale ratio is much smaller than unity. Following \cite{1994PhRvD..50.3816C}, we can define $T_{\rm orb}/T_{\rm rad}$ in terms of the tangential to radial velocity ratio, namely $T_{\rm orb}/T_{\rm rad} = |\dot{r}|/(r \Omega)$. Since this ratio increases for decreasing values of $r$, it reaches its largest value during the HDS at the location of the LCO. We show in Fig.~\ref{fig:adiabatic} the above ratio evaluated at the location of the LCO, for selected values of $\mu/M0$ and selected black-hole spin parameters. It can be seen that for the current example with $a/M = 0.9$, we have $T_{\rm orb}/T_{\rm rad}\sim 0.02$ at $r_{\rm LCO}/M=2.32$.

\begin{figure*}
\centering
\includegraphics[width=\hsize,clip]{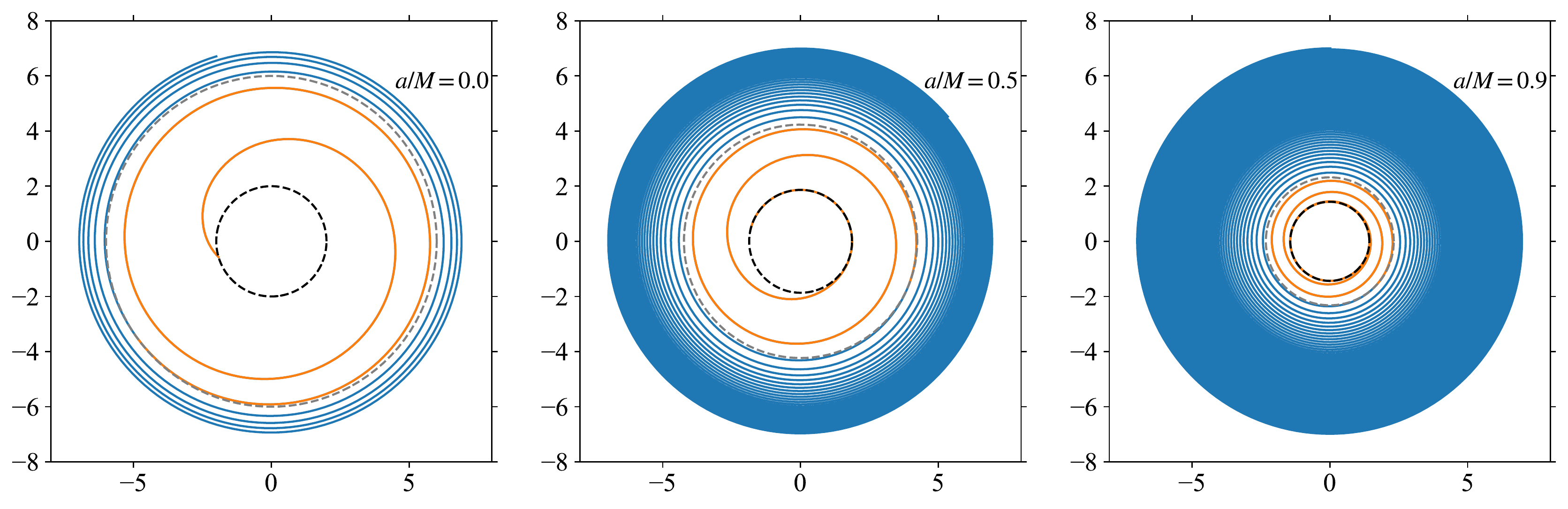}
\caption{Trajectory of a test particle in the HDS around a Schwarzschild black hole, $a/M = 0$ (left panel) and around a Kerr black hole with $a/M = 0.5$ (center panel) and with $a/M = 0.9$ (right panel). The blue part of the trajectory goes from $r = 7 M$ up to the location of the corresponding LCO (marked with a gray-dashed circle), while the orange color indicates the plunge regime. The black-hole horizon is indicated with a black-dashed circle. The mass-ratio is $\mu/M = 1/100$.}\label{fig:plunge}
\end{figure*}

Figure~\ref{fig:plunge} shows the trajectory of the test particle in the HDS around a Schwarzschild black hole, $a/M = 0$ (left panel) and around a Kerr black hole with $a/M = 0.5$  (center panel) and with $a/M = 0.9$ (right panel). In this figure the red part of the trajectory goes from $r = 7 M$ up to the location of the corresponding LCO (marked with a gray-dashed circle). The light-green color indicates the plunge regime discussed in the next section. The black-hole horizon is indicated with a black-dashed circle.

\section{Plunge into the black hole}\label{sec:6}

We consider now the evolution of the particle after reaching the LCO, namely the plunge of the particle into the black hole. A physical insight of this process can be obtained from the radial effective potential (\ref{eq:Veff}).

\begin{figure}
\centering
\includegraphics[width=\hsize,clip]{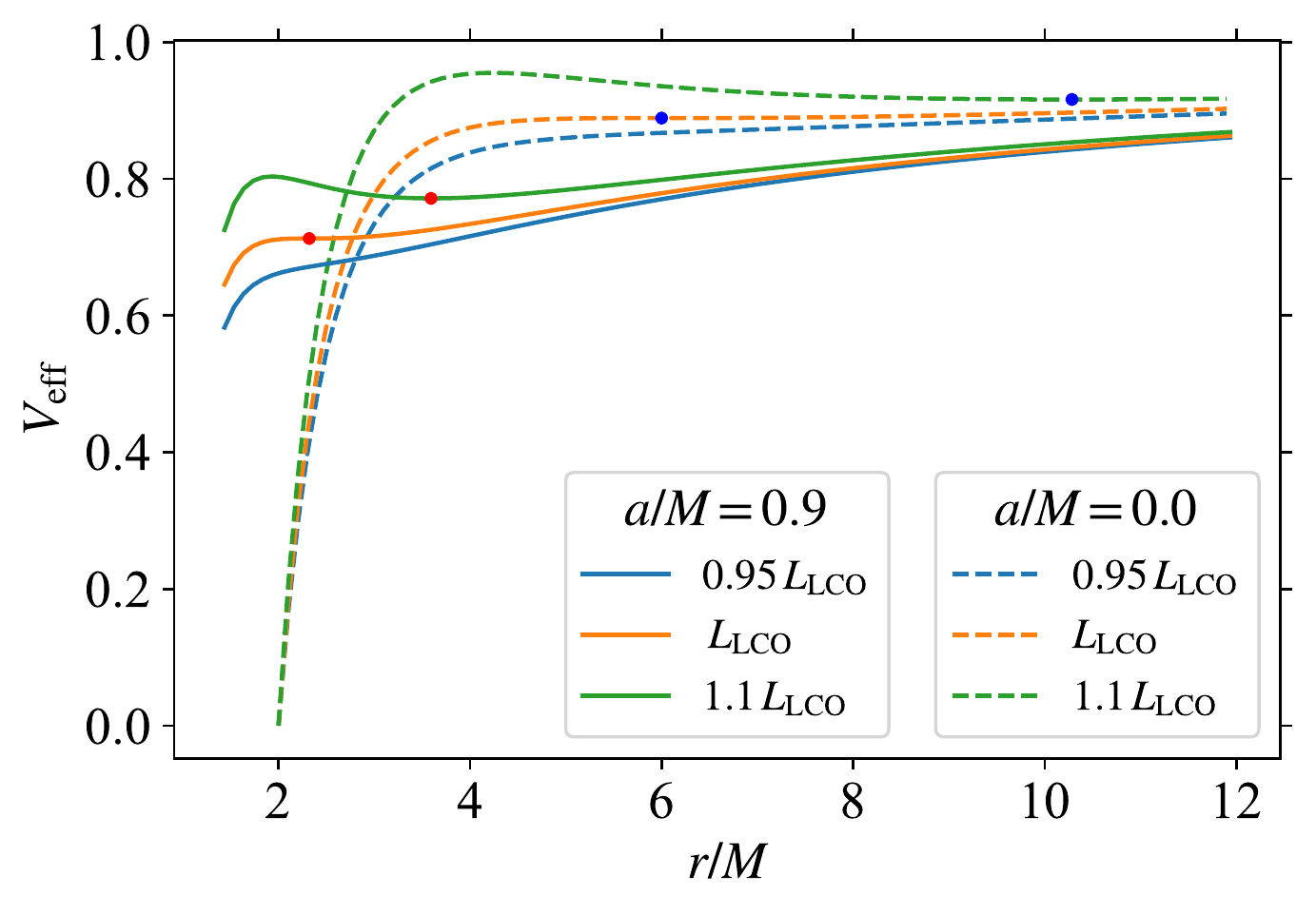}
\caption{$V_{\rm eff}$ for different values of the particle angular momentum in the case of a Schwarzschild black hole (dashed curves) and a Kerr black hole with $a/M = 0.9$ (solid curves). The blue and red dots indicate the test particle on the minimum of the potential in the Schwarzschild and Kerr cases, respectively.}\label{fig:Veff}
\end{figure}

Figure~\ref{fig:Veff} compares and contrasts $V_{\rm eff}$ as a function of $r/M$ in the case of $a/M = 0$ (Schwarzschild black hole) and $a/M = 0.9$, for three selected values of the orbital angular momentum of the particle, $L$: one larger, one equal, and one smaller than the value of the angular momentum at the LCO, $L_{\rm LCO}$. 

For $L > L_{\rm LCO}$, a small decrease in $L$ due to the gravitational-wave radiation makes the particle to change from one minimum to the next, namely it goes from a circular orbit to the next one of smaller radius. At $L = L_{\rm LCO}$ the particle reaches the LCO and, for $L < L_{\rm LCO}$, the effective potential has no minima, namely no circular orbits exist and the particle fall into the black hole.

In the realistic situation of the HDS shown in the previous section, the full numerical integration of the particle equations of motion shows that, indeed, when the particle passes the location of the LCO, it possesses a large radial momentum (see Fig.~\ref{fig:PrPphi}) and an angular momentum $L < L_{\rm LCO}$, so it continues its infall towards the black hole, smoothly, without any further radiation loss.

Fig.~\ref{fig:Veff} shows that the plunge to the black hole is markedly different for Schwarzschild and Kerr black holes: while the effective potential for $L=L_{\rm LCO}$ is zero at the horizon in the Schwarzschild case, it reaches a finite, non-zero value for the Kerr metric. The flatness of the particle effective potential from the LCO to the horizon in the case of Kerr black holes implies that very little amount of energy and angular momentum can be radiated out to infinity during the plunge from the LCO to the black-hole horizon. This can be also understood from the fact that, due to the frame-dragging effect, the particle is forced to approach the Kerr black-hole horizon tidally locked, hence it approaches the black hole with non-zero angular momentum. All these features are confirmed by our numerical integration (see below).

The plunge is geodesic thus we integrate the equations of motion in this part of its evolution in absence of energy and momentum losses, namely integrating Eqs.~(\ref{eq:rdot})--(\ref{eq:pphidot}) with ${\cal F}_r^{\rm nc} = 0$ and ${\cal F}_\phi^{\rm nc} = 0$. It is important to mention that in order to approach the horizon we use the momentum $p_{r^*}$, conjugate of the tortoise radial coordinate $r^*$ defined by $dr^*/dr = (r^2+a^2)/(r^2 - 2 M r + a^2)$, instead of the radial momentum, $p_r$.

Fig.~\ref{fig:plunge} shows the full evolution of the test particle until it reaches the black-hole horizon, for three selected cases: $a/M=0,0.5,0.9$. The blue part of the trajectory corresponds to the HDS evolution up to the passage of the particle at the location of the LCO (marked with the gray-dashed circle). The orange part of the trajectory corresponds to the final plunge into the black-hole horizon (marked with the black-dashed circle). 

\begin{figure}
\centering
\includegraphics[width=\hsize,clip]{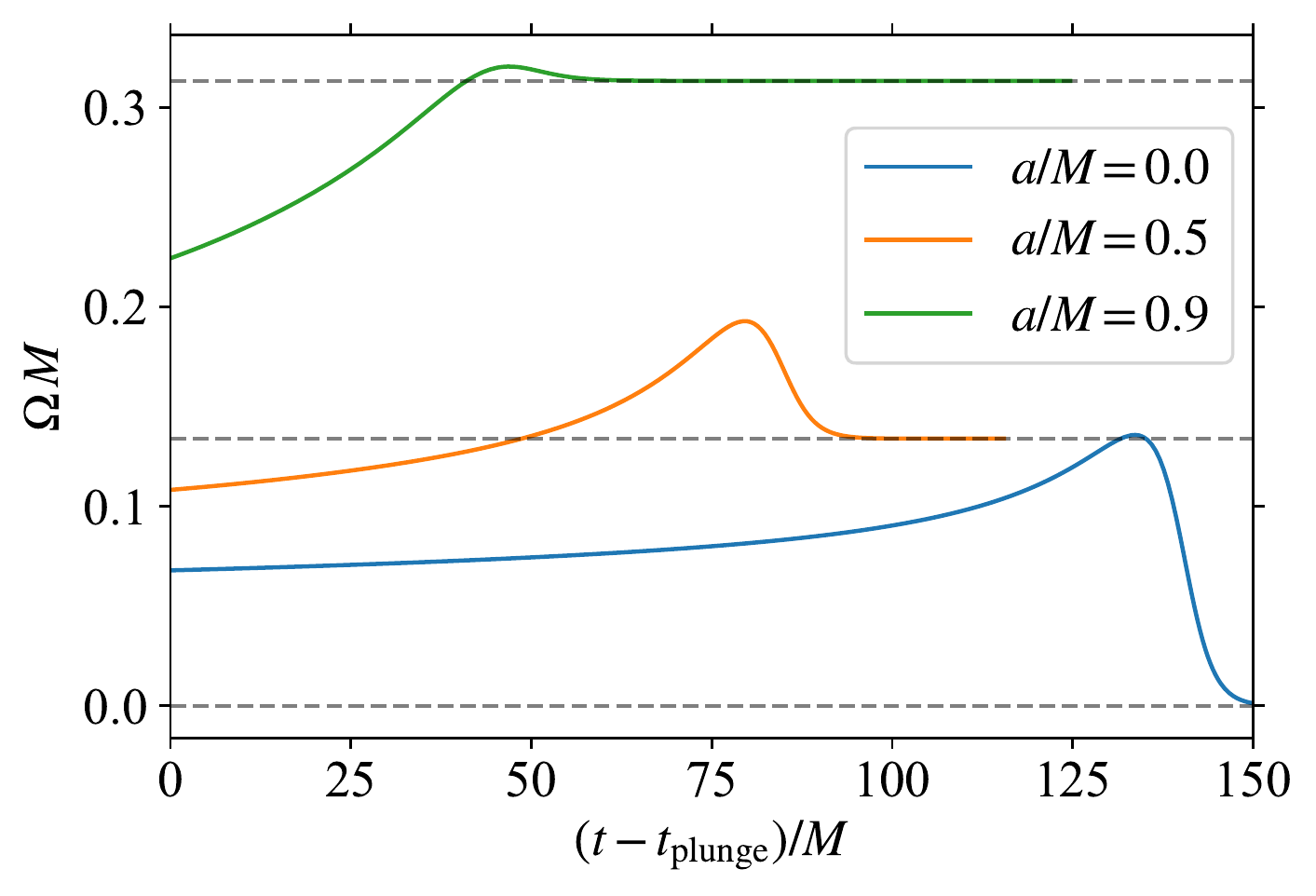}
\caption{Angular velocity of the particle, $\Omega = d\phi/dt$, during the plunge regime, namely after crossing the location of the LCO of a Kerr black with $a/M = 0.9$ (blue curve), $a/M = 0.5$ (orange curve) and in the case of Schwarzschild black hole (green curve). The mass-ratio is $\mu/M = 1/100$. The gray-dashed horizontal lines show the corresponding values of the angular velocity of the black-hole horizon. The time $t_{\rm plunge}$ is here defined as the time of the passage of the particle at the location of the LCO.}\label{fig:wplunge}
\end{figure}

We show in Fig.~\ref{fig:wplunge} the angular velocity of the particle during this phase where it corotates and, as it approaches the horizon, it approaches tidal locking to the black hole. We recall that the angular velocity of the black-hole horizon is given by $\Omega_+= a/(r_+^2 + a^2)$ \cite{1973blho.conf..451R}, where $r_+ = M + \sqrt{M^2 - a^2}$ is the black hole outer horizon radius.

\begin{figure*}
\centering
\includegraphics[width=\hsize,clip]{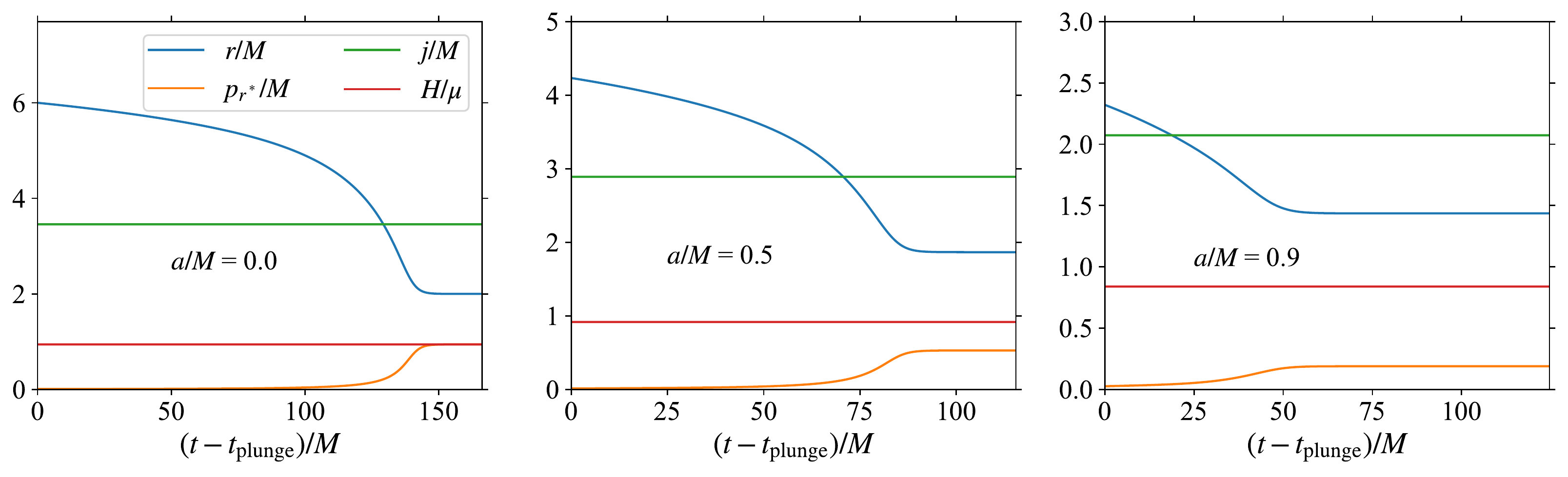}
\caption{Physical properties of a test particle in the final plunging into a Schwarzschild black hole, $a/M = 0$ (left panel), into a Kerr black hole with $a/M = 0.5$ (center panel) and into a Kerr black hole with $a/M = 0.9$ (right panel). The mass-ratio is $\mu/M = 1/100$. We show the particle radial position, $r/M$ (dotted-red), dimensionless angular momentum, $P_\phi/(\mu M) = p_\phi/M$ (dashed-blue curve), dimensionless radial momentum, $-p_{r^*}$ (dot-dashed green), and the dimensionless energy, $H/\mu$ (solid black). The time $t_{\rm plunge}$ is here defined as the time of the passage of the particle at the location of the LCO.}
\label{fig:propplunge}
\end{figure*}

Figure~\ref{fig:propplunge} shows the evolution during the final plunge phase of the energy, and the radial and the angular momentum of the particle. It can be seen that both energy and angular momentum are conserved, and the difference in effective potential between the LCO and the horizon (see Fig.~\ref{fig:Veff}) is fully converted into the particle infalling kinetic energy. We can see that for non-zero black hole rotation, in agreement with the effective potential shown in Fig.~\ref{fig:Veff}, such a difference is smaller and thus the particle approaches the horizon with lower radial velocity with respect to the Schwarzschild case, in which the particle approaches the horizon with a radial velocity approaching the speed of light. We have also included the radial position from which it can be seen the particle's approach to the horizon.

We can now compare and contrast our results with the ones obtained by different treatments of the ``transition from inspiral to plunge'' in the literature, in particular with the one in \cite{2000PhRvD..62l4022O}. They derived simplified approximate equations of motion and corresponding semi-analytic solutions by performing Taylor expansions of the energy and angular momentum of the particle around the LCO values, $E_{\rm LCO}$ and $L_{\rm LCO}$, and thus deriving an approximate effective potential. Both the angular velocity of the particle and the energy radiated in gravitational waves are assumed equal to the LCO values around the LCO. The boundary conditions are there set by imposing that the solution matches, before the LCO the adiabatic motion, and after the LCO a fully geodesic (i.e. non-radiative) plunge. With all the above they obtained semi-analytic formulas for the particle energy and angular-momentum in the final plunge phase, $E_f$ and $L_f$, there expressed as ``deficits'' with respect to the LCO values, i.e. their Eq.~(3.26) give the differences 
\begin{equation}\label{eq:deficits}
\Delta E \equiv E_f - E_{\rm LCO},\qquad \Delta L \equiv L_f-L_{\rm LCO},
\end{equation}
which are both negative. Namely they evaluate thorough the above differences the deviation of the particle's motion from the strict circular adiabatic one. Although we agree with their general qualitative picture, it can be checked using their Eq.~(3.26) and Table I, that their above ``deficits'' are much bigger than the ones we obtain by the full numerical integration of the HDS equations of motion.

\begin{figure}
\includegraphics[width=0.45\textwidth]{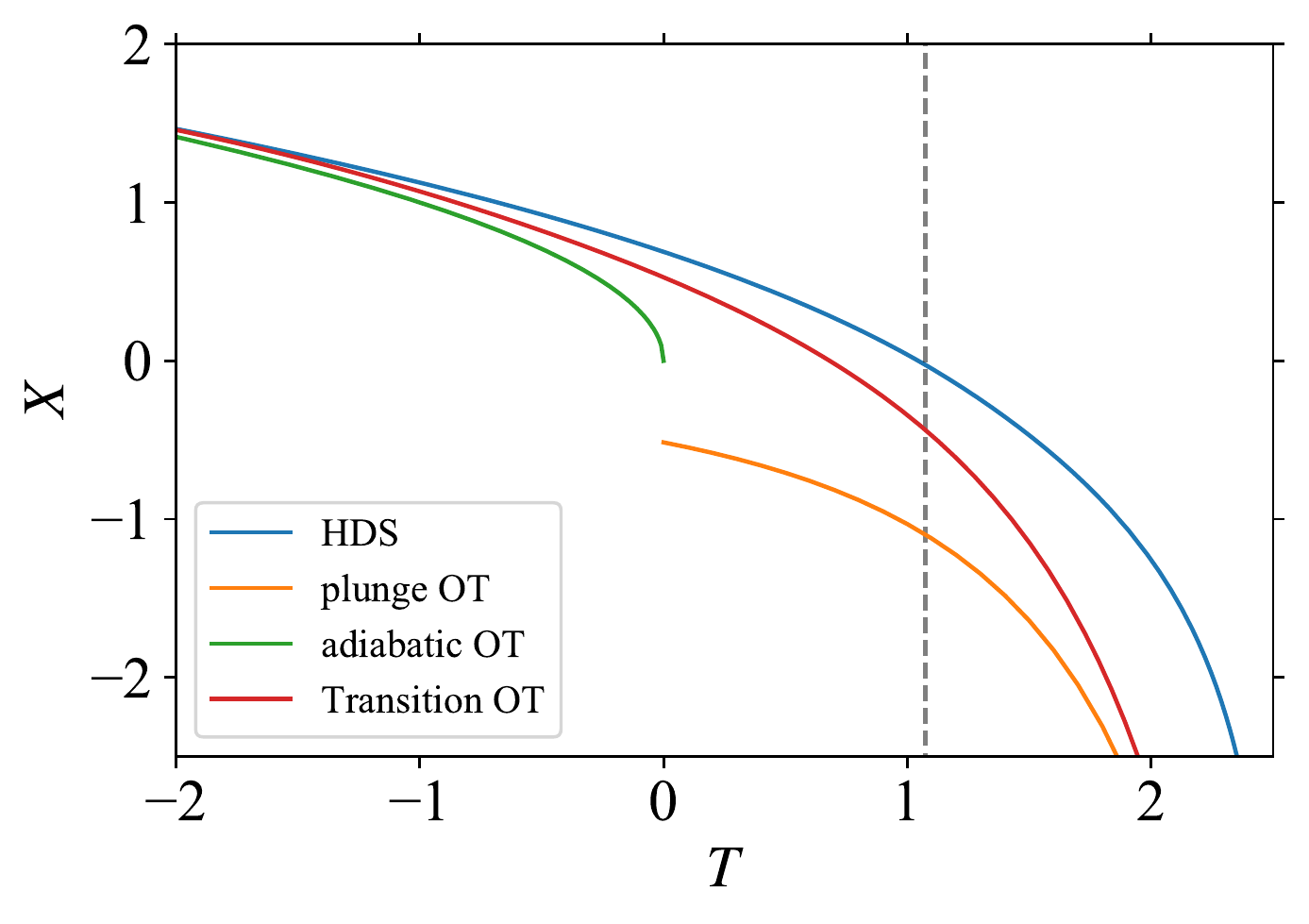}
\caption{Comparison between the results of the transition from inspiral to plunge in \cite{2000PhRvD..62l4022O} and in our present HDS approach, in the case of a particle falling into a Kerr black hole with spin parameter $a/M = 0.9$. The variable $X$ is normalized difference between the radial (Boyer-Lindquist) position of the particle  and the LCO. The variable $T$ is the normalized proper time shifted so that $T=0$ when $r=r_{\textrm{LCO}}$ for the adiabatic trajectory.
The details on the normalization can be found in \cite{2000PhRvD..62l4022O}. It can be seen that near the LCO, i.e. around $X = 0$ and $T < 0$, the two trajectories are similar, but the farther from $T > 0$ the more separated they become. Clearly, this is consistent with the expansion of the effective potential around the LCO which is expected to be valid only near $X=T=0$. The vertical dashed line corresponds to $t_{\rm plunge}$ in our present HDS approach, which in these units is $T\approx 1$.}
\label{fig:OT}
\end{figure}

{Figure~\ref{fig:OT} compares and contrasts, for the case of a particle falling into a Kerr black hole with spin parameter $a/M = 0.9$, the particle radial trajectory, near the location of the LCO, derived from the treatment in \cite{2000PhRvD..62l4022O} with the one of our present HDS approach. It can be seen that the two solutions converge at a large distance from the black hole. This indicates, as we have explained in Sec.~\ref{sec:5}, that these two solutions satisfy the same initial condition set by the adiabatic approximation.}

\begin{figure}
\centering
\includegraphics[width=\hsize,clip]{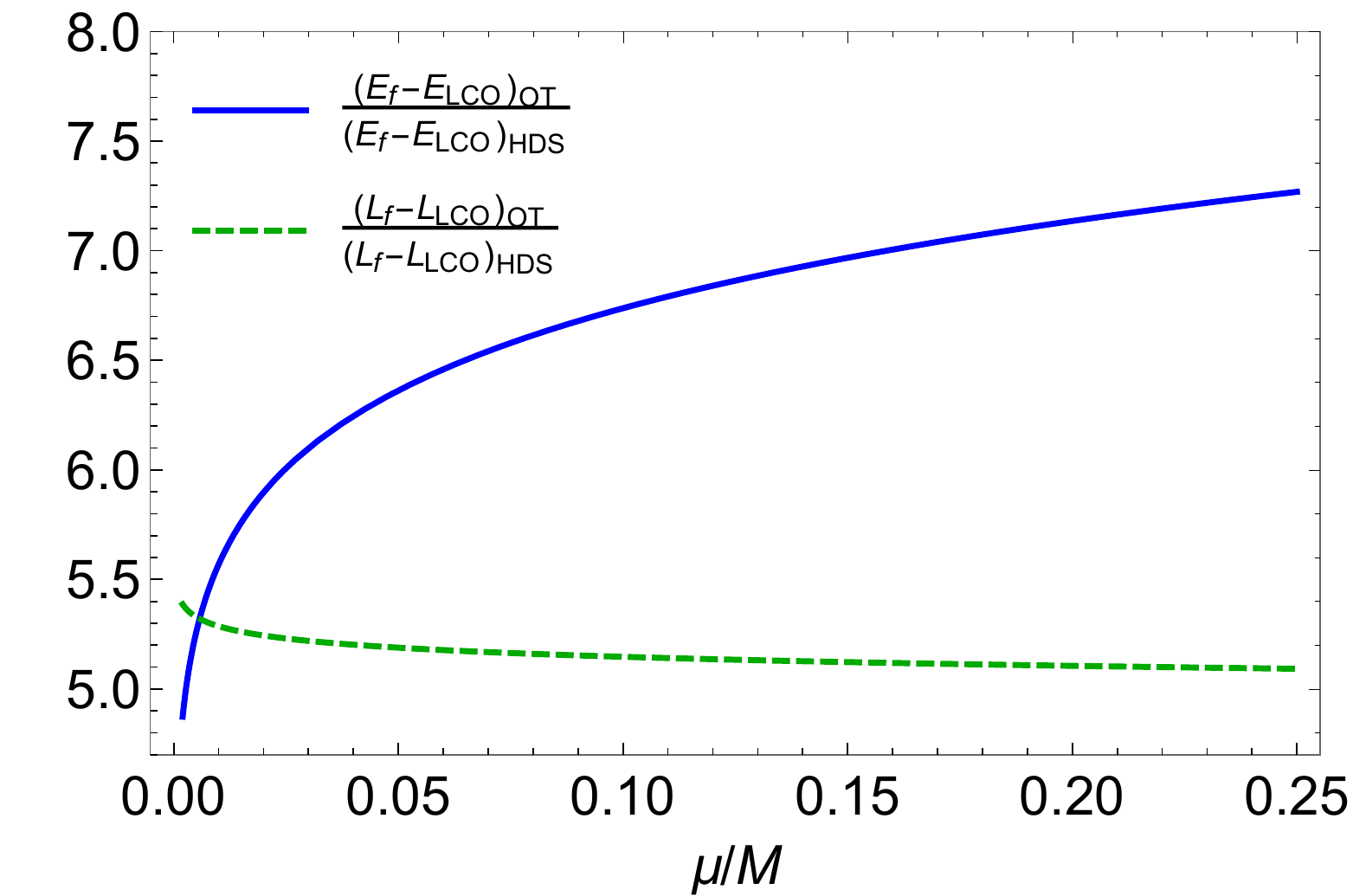}
\caption{Comparison and contrast, for the case $a/M = 0.9$ and as a function of the mass-ratio $\mu/M$, of the quantities $E_f-E_{\rm LCO}$ and $L_f-L_{\rm LCO}$ obtained from our HDS, i.e. with $E_f \equiv H(t_{\rm plunge})$ and $L_f = p_\phi(t_{\rm plunge})$, with respect to the semi-analytic formulas (3.26) in Ref.~\cite{2000PhRvD..62l4022O} (indicated with the subscript OT). We define the time $t_{\rm plunge}$ as the instant in which there is the passage of the test particle at the location of the LCO. We recall that for $a/M = 0.9$, we have $E_{\rm LCO}/\mu = 0.84425$ and $L_{\rm LCO}/(\mu M) = 2.09978$. See text for further details.}\label{fig:comparisonOT}
\end{figure}

We show in Fig.~\ref{fig:comparisonOT} a comparison, for the case $a/M = 0.9$ and as a function of the mass-ratio $\mu/M$, of the energy and angular momentum ``deficits'' $\Delta E$ and $\Delta L$ obtained from our HDS, i.e. adopting $E_f \equiv H(t_{\rm plunge})$ and $L_f = p_\phi(t_{\rm plunge})$, with the ones given by semi-analytic formulas (3.26) of Ref.~\cite{2000PhRvD..62l4022O}. We recall that we define the time $t_{\rm plunge}$ as the time at which there is the passage of the test particle at the location of the LCO. We recall that for $a/M = 0.9$, the energy and angular momentum of the LCO are, respectively, $E_{\rm LCO}/\mu = 0.84425$ and $L_{\rm LCO}/(\mu M) = 2.09978$.

The above implies a larger amount of gravitational radiation in the treatment of Ref.~\cite{2000PhRvD..62l4022O}. Possibly, the main reason for this additional gravitational-wave emission is their assumption of keeping the particle radiating waves at the rate of the LCO (which is the highest one of all circular orbits; see Fig.~\ref{fig:dEdt}), before and well after crossing it. We found that extrapolation of such an approximation much beyond the LCO is what causes a larger discrepancy with our results. {It is clear that such extra radiation is needed under the assumption of strict circular orbits since, in absence of such a radiation, there is no reason for the particle to plunge into the black hole because the LCO is a stable orbit. When the radial drift is considered, as in the HDS, there is an increasing contribution of the radial momentum (see Figs.~\ref{fig:PrPphi} and \ref{fig:propplunge}) {that modifies the particle's effective potential. In this context} the LCO does not play any special role and only assists to the passage of the plunging particle in view of the acquired radial momentum. This can be also appreciated from Fig.~\ref{fig:OT}. As a result, we expect that in the HDS the test particle smoothly falls into the Kerr black hole with a negligible gravitational-wave emission. The only radiation comes from the non-circular plunge trajectory followed by the particle (e.g. the orange-color trajectories in Fig.~\ref{fig:plunge}) which must be much smaller than the one emitted by the test particle per orbit at the LCO. This explains the additional gravitational-wave radiation obtained in \cite{2000PhRvD..62l4022O} with respect to our results.}

{We turn now to evaluate the consequences of the differences in the amount of gravitational radiation emitted in the in the estimate of the mass of the final black hole. The latter can be estimated as 
\begin{equation}\label{eq:Mbh}
M_f = M + \Delta E_{\rm rad},\qquad \Delta E_{\rm rad} \equiv E_f - \mu < 0,
\end{equation}
where $M$ is the total mass of the merging black-hole binary and $E_f$ the final energy of the test particle. Thus, one can write the energy radiated as
\begin{equation}\label{eq:DeltaE}
\Delta E_{\rm rad} = \Delta E_{\rm ad} + \Delta E,
\end{equation}
where $\Delta E_{\rm ad} \equiv E_{\rm LCO}- \mu$ is the energy radiated up to the location of the LCO within the adiabatic, strict circular motion approximation, and $\Delta E$ is the energy ``deficit'' with respect to such an approximation as defined in Eq.~\eqref{eq:deficits}. It can be checked that, for instance in the case of an equal-mass binary, the contribution of $\Delta E$ to $M_f$ (in the $a/M = 0.9$ case under examination) is about 10\%, while in our HDS case it is only about 1\% (see Fig.~\ref{fig:comparisonOT}).}

{This additional amount of gravitational radiation of the treatment in \cite{2000PhRvD..62l4022O} will lead to a more energetic gravitational waveform in the final merging phase. It is interesting that a similarly energetic plunge leading to a burst of radiation in the black-hole binary merger appears in the numerical-relativity waveforms of the SXS catalog \cite{SXS:catalog}. Such a feature is also found in the binary mergers modeled via the EOB formalism \cite{1999PhRvD..59h4006B} which adopts a treatment as the one in \cite{2000PhRvD..62l4022O} for the plunge phase (see Ref.~\cite{2000PhRvD..62f4015B} for details). Therefore, the above discrepancy between the HDS treatment and the one of Ref.~\cite{2000PhRvD..62l4022O} acquires relevance in view of the large use, by the LIGO-Virgo Collaboration, of the above waveforms with such a burst structure in the merger for the binary parameter estimation (see e.g. \cite{2016PhRvL.116f1102A} for the case of the event GW150914). In this line it is most important to recall the recent result of an independent analysis of the GW150914 event that shows the incompatibility of the LIGO-Virgo data with the presence of the aforementioned gravitational-radiation burst, in clear contrast with the currently used waveform templates (see Figs.~5 and 10 in Ref.~\cite{2018JCAP...02..013L} and also Refs.~\cite{2016JCAP...08..029N,2016JCAP...10..014L,2017JCAP...08..013C}, for further details).}

{We assess now the dependence of the radiated energy and angular momentum on the mass-ratio, $\mu/M$. For a given a black-hole spin parameter, the semi-analytic treatment of Ref.~\cite{2000PhRvD..62l4022O} predicts that the above-defined energy and angular momentum ``deficits'' scale with the mass-ratio as $|\Delta E|_{\rm OT}/\mu\propto (\mu/M)^{4/5}$ and $|\Delta L|_{\rm OT}/(\mu M) \propto (\mu/M)^{4/5}$. We find from our numerical computations that, approximately, $\Delta E_{\rm HDS}/\mu\propto (\mu/M)^{0.72}$ and $\Delta L_{\rm HDS}/(\mu M)\propto (\mu/M)^{0.81}$. The different scaling with the mass-ratio for the energy and angular momentum radiated within the HDS treatment implies that the ratio $\Delta E/\Delta L$ depends on the mass-ratio. This is consistent with the fact that the particle in the HDS case does not follow strict circular orbits, as in the case of Ref.~\cite{2000PhRvD..62l4022O}. For strict circular orbits, the energy to angular momentum ratio gives, at any radius, the value of the particle's angular velocity which depends only on the black-hole spin. Figure~\ref{fig:comparisonOT} shows explicitly these differences.}

\begin{figure}
\centering
\includegraphics[width=\hsize,clip]{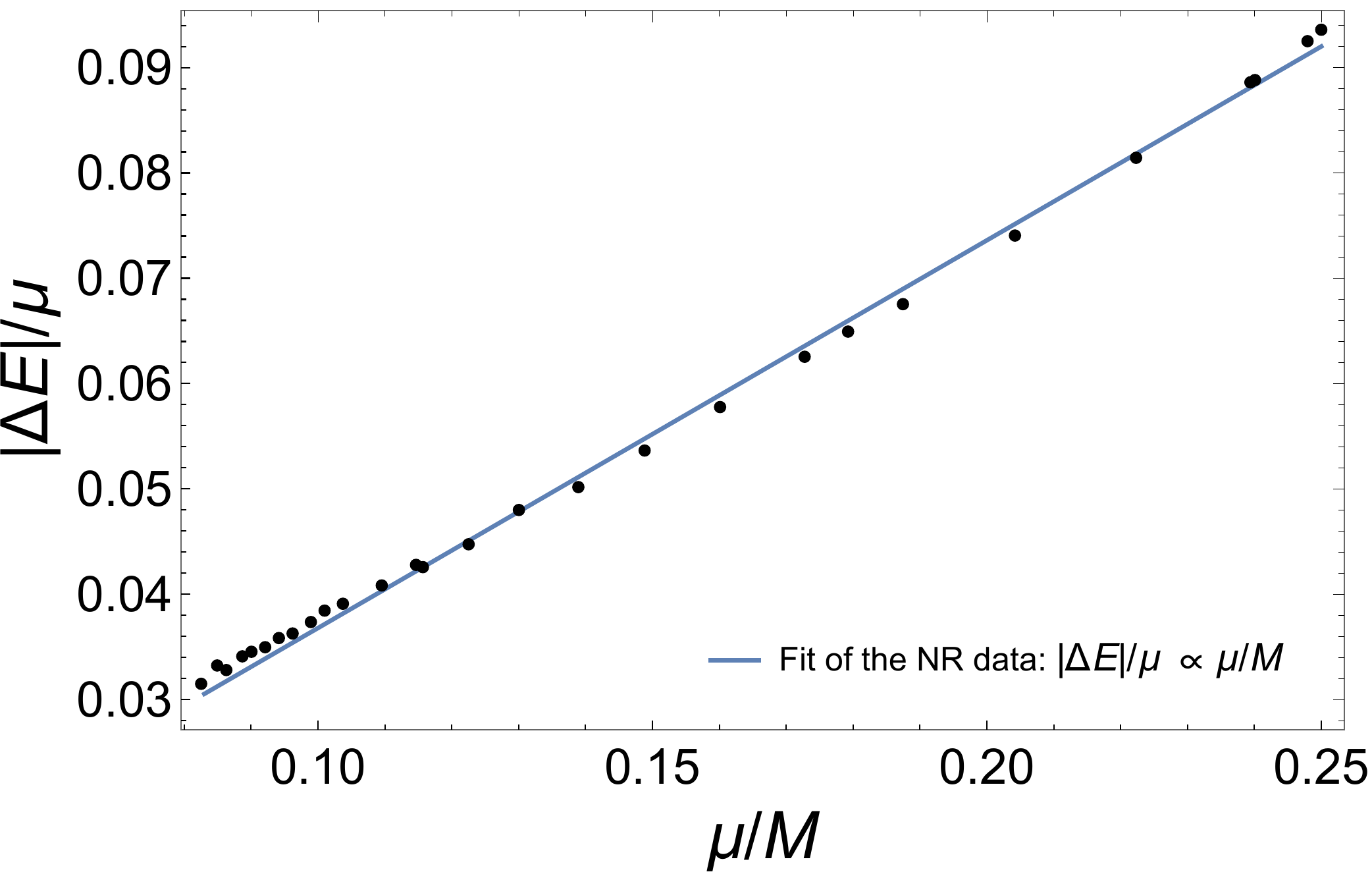}
\caption{{Energy ``deficit'' $\Delta E/\mu$ estimated from the numerical-relativity simulations of the SXS catalog \cite{SXS:catalog}. See text for details. It is surprising that it scales in the same way as the energy radiated in gravitational waves in the case of a purely radial plunge of a test particle into a Schwarzschild black hole, i.e.~$\Delta E/\mu \propto \mu/M$ \cite{1971PhRvL..27.1466D} (see also Sec.~\ref{sec:2}).}}\label{fig:DeltaENR}
\end{figure}

{It is interesting at this point to compare and contrast the energy radiated predicted by the above models with the one declared in numerical-relativity simulations. Such a comparison can be done only in the comparable mass regime since no numerical-relativity simulations are available for small values of the mass-ratio. We recall that a qualitative and quantitative comparison in the comparable mass regime of the HDS and the numerical-relativity waveforms has been recently performed in \cite{2018JCAP...02..030R} finding an expected agreement between the two treatments. We thus proceed further here adopting the working hypothesis that the test particle treatment might be a good approximation of the real two-body system of comparable masses. First, we compute from the data available in the SXS catalog \cite{SXS:catalog} the corresponding energy ``deficit'' $\Delta E/\mu$, for different mass-ratios $\mu/M$. We do this following Eq.~\eqref{eq:Mbh}, namely we subtract off the value of $\Delta E_{\rm ad}$, which depends only on the black hole spin, to the mass of the final Kerr black hole obtained in the simulation. Figure~\ref{fig:DeltaENR} shows the results for $\mu/M= 0.08$--0.25. It can be seen that $\Delta E/\mu$ obtained from the numerical-relativity data scales linearly with $\mu/M$. It is quite surprising that these simulations follow exactly the same scaling of the energy radiated in gravitational waves in the case of a purely radial plunge of a test particle into a Schwarzschild black hole, i.e.~$\Delta E/\mu \propto \mu/M$ \cite{1971PhRvL..27.1466D} (see also Sec.~\ref{sec:2}), and not the one expected from the particle plunge derived either in \cite{2000PhRvD..62l4022O} or in the present work.}

\section{Conclusions}\label{sec:7}

Two different cases of emission of gravitational waves in the strong-field limit from test particles falling into an already formed black hole have been considered. 
We have both reviewed the historical key contributions and our current results, compared and contrasted with the ones in the literature. 
The first case is the one of a particle starting from infinite distance and plunging either initially at rest or with finite amount of kinetic energy. This case leads necessarily to an emission of a burst of gravitational waves always composed of three different components, ``\emph{a precursor, a main burst and a ringing tail}''. The amplitude of the burst is a function of the angular momentum of the particle and of the black hole and the case of a Schwarzschild versus the Kerr black hole introduces only quantitative but not qualitative differences. The structures of the multipoles can be in principle used to determine the angular momentum both of the particle and of the black hole. 

The second case of a particle starting from a finite radius in a circular orbit leads to a novel HDS leading finally to a smooth merging into the black hole. The difference between the Schwarzschild and the Kerr metric is specially manifest at the LCO and after when no radiation reaction is taken into account. 
When the radiation reaction is taken into account, as evidenced in Fig.~\ref{fig:plunge}, the clear appearance of the HDS occurs with an increasing contribution of the radial momentum as the particle approaches the horizon (see Figs.~\ref{fig:PrPphi} and \ref{fig:propplunge}). This phenomenon is further enhanced by the essentially different effective potential between the Schwarzschild and Kerr cases. The LCO in this context does not witness any special role and only assists to the passage of the plunging particle in view of the acquired radial momentum in its previous history. The final result is the one of a test particle smoothly merging in the Kerr black hole without any burst. We have compared and contrasted our treatment with the approximate one in Ref.~\cite{2000PhRvD..62l4022O} and found that in the latter a larger amount of gravitational radiation is emitted in the final transition to the plunge phase. We have shown the effect that such a difference may cause in the estimation of the mass of the final black hole formed in the merger of a binary black hole.

{We have derived, within the HDS treatment, how the energy and the angular momentum radiated in gravitational waves scale with the mass-ratio $\mu/M$. We have shown the difference of such scaling law with the results of Ref.~\cite{2000PhRvD..62l4022O}. Stimulated by the unexpected agreement of the HDS and numerical-relativity waveforms in the comparable mass regime recently found in \cite{2018JCAP...02..030R}, we proceeded here to make a comparison of the energy radiated and its scaling with the mass-ratio in the test particle approximation with the one obtained in numerical-relativity simulations. To do this we use the SXS catalog simulations \cite{SXS:catalog} in the mass-ratio range $\mu/M= 0.08$--0.25. The results are shown in Fig.~\ref{fig:DeltaENR}. Surprisingly, we found that $\Delta E/\mu$ in the numerical-relativity simulations scales linearly with $\mu/M$, namely it follows exactly the same scaling of a purely radial plunge of a test particle into a Schwarzschild black hole \cite{1971PhRvL..27.1466D}, but not the scaling predicted in \cite{2000PhRvD..62l4022O} nor the one of the HDS treatment presented here for a particle inspiraling and plunging into a Kerr black hole.}

{Besides the theoretical interest, the above results are relevant in view of the large use, by the LIGO-Virgo Collaboration (see e.g. the case of GW150914 in \cite{2016PhRvL.116f1102A}), of waveforms based on the numerical-relativity simulations \cite{SXS:catalog} and on the EOB formalism \cite{1999PhRvD..59h4006B} which adopts a plunge phase following the treatment of Ref.~\cite{2000PhRvD..62f4015B}. Along these lines, it is also important to recall a most important recent result of an independent analysis of GW150914 that shows the incompatibility of the LIGO-Virgo data with the presence of such a burst, in clear contrast with the currently used waveform templates (see Figs.~5 and 10 in Ref.~\cite{2018JCAP...02..013L} and see also Refs.~\cite{2016JCAP...08..029N,2016JCAP...10..014L,2017JCAP...08..013C}).}

\acknowledgements

We had the pleasure during the entire development of this work to discuss in our ICRANet headquarter in Pescara with Prof. Roy Kerr. To him, to the entire ICRANet Faculty and Staff goes our gratitude. {We thank the Referee for the comments and suggestions that helped to improve the presentation of our results.}
  

\begin{thebibliography}{65}
\expandafter\ifx\csname natexlab\endcsname\relax\def\natexlab#1{#1}\fi
\expandafter\ifx\csname bibnamefont\endcsname\relax
  \def\bibnamefont#1{#1}\fi
\expandafter\ifx\csname bibfnamefont\endcsname\relax
  \def\bibfnamefont#1{#1}\fi
\expandafter\ifx\csname citenamefont\endcsname\relax
  \def\citenamefont#1{#1}\fi
\expandafter\ifx\csname url\endcsname\relax
  \def\url#1{\texttt{#1}}\fi
\expandafter\ifx\csname urlprefix\endcsname\relax\def\urlprefix{URL }\fi
\providecommand{\bibinfo}[2]{#2}
\providecommand{\eprint}[2][]{\url{#2}}

\bibitem[{\citenamefont{{Schmidt}}(1963)}]{1963Natur.197.1040S}
\bibinfo{author}{\bibfnamefont{M.}~\bibnamefont{{Schmidt}}},
  \bibinfo{journal}{\nat} \textbf{\bibinfo{volume}{197}}, \bibinfo{pages}{1040}
  (\bibinfo{year}{1963}).

\bibitem[{\citenamefont{{Hewish} et~al.}(1968)\citenamefont{{Hewish}, {Bell},
  {Pilkington}, {Scott}, and {Collins}}}]{1968Natur.217..709H}
\bibinfo{author}{\bibfnamefont{A.}~\bibnamefont{{Hewish}}},
  \bibinfo{author}{\bibfnamefont{S.~J.} \bibnamefont{{Bell}}},
  \bibinfo{author}{\bibfnamefont{J.~D.~H.} \bibnamefont{{Pilkington}}},
  \bibinfo{author}{\bibfnamefont{P.~F.} \bibnamefont{{Scott}}},
  \bibnamefont{and} \bibinfo{author}{\bibfnamefont{R.~A.}
  \bibnamefont{{Collins}}}, \bibinfo{journal}{\nat}
  \textbf{\bibinfo{volume}{217}}, \bibinfo{pages}{709} (\bibinfo{year}{1968}).

\bibitem[{\citenamefont{{Giacconi} et~al.}(1972)\citenamefont{{Giacconi},
  {Murray}, {Gursky}, {Kellogg}, {Schreier}, and
  {Tananbaum}}}]{1972ApJ...178..281G}
\bibinfo{author}{\bibfnamefont{R.}~\bibnamefont{{Giacconi}}},
  \bibinfo{author}{\bibfnamefont{S.}~\bibnamefont{{Murray}}},
  \bibinfo{author}{\bibfnamefont{H.}~\bibnamefont{{Gursky}}},
  \bibinfo{author}{\bibfnamefont{E.}~\bibnamefont{{Kellogg}}},
  \bibinfo{author}{\bibfnamefont{E.}~\bibnamefont{{Schreier}}},
  \bibnamefont{and}
  \bibinfo{author}{\bibfnamefont{H.}~\bibnamefont{{Tananbaum}}},
  \bibinfo{journal}{\apj} \textbf{\bibinfo{volume}{178}}, \bibinfo{pages}{281}
  (\bibinfo{year}{1972}).

\bibitem[{\citenamefont{{Leach} and {Ruffini}}(1973)}]{1973ApJ...180L..15L}
\bibinfo{author}{\bibfnamefont{R.~W.} \bibnamefont{{Leach}}} \bibnamefont{and}
  \bibinfo{author}{\bibfnamefont{R.}~\bibnamefont{{Ruffini}}},
  \bibinfo{journal}{\apj} \textbf{\bibinfo{volume}{180}}, \bibinfo{pages}{L15}
  (\bibinfo{year}{1973}).

\bibitem[{\citenamefont{{Giacconi} and {Ruffini}}(1978)}]{1978pans.proc.....G}
\bibinfo{editor}{\bibfnamefont{R.}~\bibnamefont{{Giacconi}}} \bibnamefont{and}
  \bibinfo{editor}{\bibfnamefont{R.}~\bibnamefont{{Ruffini}}}, eds.,
  \emph{\bibinfo{title}{{Physics and astrophysics of neutron stars and black
  holes}}} (\bibinfo{year}{1978}).

\bibitem[{\citenamefont{{Hulse} and {Taylor}}(1975)}]{1975ApJ...195L..51H}
\bibinfo{author}{\bibfnamefont{R.~A.} \bibnamefont{{Hulse}}} \bibnamefont{and}
  \bibinfo{author}{\bibfnamefont{J.~H.} \bibnamefont{{Taylor}}},
  \bibinfo{journal}{\apj} \textbf{\bibinfo{volume}{195}}, \bibinfo{pages}{L51}
  (\bibinfo{year}{1975}).

\bibitem[{\citenamefont{{Weisberg} and {Taylor}}(2005)}]{2005ASPC..328...25W}
\bibinfo{author}{\bibfnamefont{J.~M.} \bibnamefont{{Weisberg}}}
  \bibnamefont{and} \bibinfo{author}{\bibfnamefont{J.~H.}
  \bibnamefont{{Taylor}}}, in \emph{\bibinfo{booktitle}{Binary Radio Pulsars}},
  edited by \bibinfo{editor}{\bibfnamefont{F.~A.} \bibnamefont{{Rasio}}}
  \bibnamefont{and} \bibinfo{editor}{\bibfnamefont{I.~H.}
  \bibnamefont{{Stairs}}} (\bibinfo{year}{2005}), vol. \bibinfo{volume}{328} of
  \emph{\bibinfo{series}{Astronomical Society of the Pacific Conference
  Series}}, p.~\bibinfo{pages}{25}, \eprint{astro-ph/0407149}.

\bibitem[{\citenamefont{{Gursky} and {Ruffini}}(1975)}]{1975ASSL...48.....G}
\bibinfo{editor}{\bibfnamefont{H.}~\bibnamefont{{Gursky}}} \bibnamefont{and}
  \bibinfo{editor}{\bibfnamefont{R.}~\bibnamefont{{Ruffini}}}, eds.,
  \emph{\bibinfo{title}{{Neutron stars, black holes and binary X-ray sources;
  Proceedings of the Annual Meeting, San Francisco, Calif., February 28,
  1974}}}, vol.~\bibinfo{volume}{48} of \emph{\bibinfo{series}{Astrophysics and
  Space Science Library}} (\bibinfo{year}{1975}).

\bibitem[{\citenamefont{{Abbott}
  et~al.}(2016{\natexlab{a}})\citenamefont{{Abbott}, {Abbott}, {Abbott},
  {Abernathy}, {Acernese}, {Ackley}, {Adams}, {Adams}, {Addesso}, {Adhikari}
  et~al.}}]{2016PhRvL.116f1102A}
\bibinfo{author}{\bibfnamefont{B.~P.} \bibnamefont{{Abbott}}},
  \bibinfo{author}{\bibfnamefont{R.}~\bibnamefont{{Abbott}}},
  \bibinfo{author}{\bibfnamefont{T.~D.} \bibnamefont{{Abbott}}},
  \bibinfo{author}{\bibfnamefont{M.~R.} \bibnamefont{{Abernathy}}},
  \bibinfo{author}{\bibfnamefont{F.}~\bibnamefont{{Acernese}}},
  \bibinfo{author}{\bibfnamefont{K.}~\bibnamefont{{Ackley}}},
  \bibinfo{author}{\bibfnamefont{C.}~\bibnamefont{{Adams}}},
  \bibinfo{author}{\bibfnamefont{T.}~\bibnamefont{{Adams}}},
  \bibinfo{author}{\bibfnamefont{P.}~\bibnamefont{{Addesso}}},
  \bibinfo{author}{\bibfnamefont{R.~X.} \bibnamefont{{Adhikari}}},
  \bibnamefont{et~al.}, \bibinfo{journal}{\prl} \textbf{\bibinfo{volume}{116}},
  \bibinfo{eid}{061102} (\bibinfo{year}{2016}{\natexlab{a}}),
  \eprint{1602.03837}.

\bibitem[{\citenamefont{{Abbott}
  et~al.}(2016{\natexlab{b}})\citenamefont{{Abbott}, {Abbott}, {Abbott},
  {Abernathy}, {Acernese}, {Ackley}, {Adams}, {Adams}, {Addesso}, {Adhikari}
  et~al.}}]{2016PhRvL.116x1103A}
\bibinfo{author}{\bibfnamefont{B.~P.} \bibnamefont{{Abbott}}},
  \bibinfo{author}{\bibfnamefont{R.}~\bibnamefont{{Abbott}}},
  \bibinfo{author}{\bibfnamefont{T.~D.} \bibnamefont{{Abbott}}},
  \bibinfo{author}{\bibfnamefont{M.~R.} \bibnamefont{{Abernathy}}},
  \bibinfo{author}{\bibfnamefont{F.}~\bibnamefont{{Acernese}}},
  \bibinfo{author}{\bibfnamefont{K.}~\bibnamefont{{Ackley}}},
  \bibinfo{author}{\bibfnamefont{C.}~\bibnamefont{{Adams}}},
  \bibinfo{author}{\bibfnamefont{T.}~\bibnamefont{{Adams}}},
  \bibinfo{author}{\bibfnamefont{P.}~\bibnamefont{{Addesso}}},
  \bibinfo{author}{\bibfnamefont{R.~X.} \bibnamefont{{Adhikari}}},
  \bibnamefont{et~al.}, \bibinfo{journal}{Physical Review Letters}
  \textbf{\bibinfo{volume}{116}}, \bibinfo{eid}{241103}
  (\bibinfo{year}{2016}{\natexlab{b}}), \eprint{1606.04855}.

\bibitem[{\citenamefont{{Abbott} et~al.}(2017)\citenamefont{{Abbott}, {Abbott},
  {Abbott}, {Acernese}, {Ackley}, {Adams}, {Adams}, {Addesso}, {Adhikari},
  {Adya} et~al.}}]{2017PhRvL.118v1101A}
\bibinfo{author}{\bibfnamefont{B.~P.} \bibnamefont{{Abbott}}},
  \bibinfo{author}{\bibfnamefont{R.}~\bibnamefont{{Abbott}}},
  \bibinfo{author}{\bibfnamefont{T.~D.} \bibnamefont{{Abbott}}},
  \bibinfo{author}{\bibfnamefont{F.}~\bibnamefont{{Acernese}}},
  \bibinfo{author}{\bibfnamefont{K.}~\bibnamefont{{Ackley}}},
  \bibinfo{author}{\bibfnamefont{C.}~\bibnamefont{{Adams}}},
  \bibinfo{author}{\bibfnamefont{T.}~\bibnamefont{{Adams}}},
  \bibinfo{author}{\bibfnamefont{P.}~\bibnamefont{{Addesso}}},
  \bibinfo{author}{\bibfnamefont{R.~X.} \bibnamefont{{Adhikari}}},
  \bibinfo{author}{\bibfnamefont{V.~B.} \bibnamefont{{Adya}}},
  \bibnamefont{et~al.}, \bibinfo{journal}{Physical Review Letters}
  \textbf{\bibinfo{volume}{118}}, \bibinfo{eid}{221101} (\bibinfo{year}{2017}),
  \eprint{1706.01812}.

\bibitem[{\citenamefont{{Ruffini} and {Wheeler}}(1969)}]{1971ESRSP..52...45R}
\bibinfo{author}{\bibfnamefont{R.}~\bibnamefont{{Ruffini}}} \bibnamefont{and}
  \bibinfo{author}{\bibfnamefont{J.~A.} \bibnamefont{{Wheeler}}},
  \bibinfo{journal}{ESRO, SP, No.~52, p.~45 - 174}
  \textbf{\bibinfo{volume}{52}}, \bibinfo{pages}{45} (\bibinfo{year}{1969}).

\bibitem[{\citenamefont{{Landau} and {Lifshitz}}(1975)}]{1975ctf..book.....L}
\bibinfo{author}{\bibfnamefont{L.~D.} \bibnamefont{{Landau}}} \bibnamefont{and}
  \bibinfo{author}{\bibfnamefont{E.~M.} \bibnamefont{{Lifshitz}}},
  \emph{\bibinfo{title}{{The classical theory of fields, 4th rev.engl.ed.}}}
  (\bibinfo{publisher}{Oxford: Pergamon Press}, \bibinfo{year}{1975}).

\bibitem[{SXS()}]{SXS:catalog}
\bibinfo{howpublished}{\url{http://www.black-holes.org/waveforms}}.

\bibitem[{\citenamefont{{Rodr{\'{\i}}guez}
  et~al.}(2018)\citenamefont{{Rodr{\'{\i}}guez}, {Rueda}, and
  {Ruffini}}}]{2018JCAP...02..030R}
\bibinfo{author}{\bibfnamefont{J.~F.} \bibnamefont{{Rodr{\'{\i}}guez}}},
  \bibinfo{author}{\bibfnamefont{J.~A.} \bibnamefont{{Rueda}}},
  \bibnamefont{and}
  \bibinfo{author}{\bibfnamefont{R.}~\bibnamefont{{Ruffini}}},
  \bibinfo{journal}{\jcap} \textbf{\bibinfo{volume}{2}}, \bibinfo{eid}{030}
  (\bibinfo{year}{2018}), \eprint{1706.07704}.

\bibitem[{\citenamefont{{Ori} and {Thorne}}(2000)}]{2000PhRvD..62l4022O}
\bibinfo{author}{\bibfnamefont{A.}~\bibnamefont{{Ori}}} \bibnamefont{and}
  \bibinfo{author}{\bibfnamefont{K.~S.} \bibnamefont{{Thorne}}},
  \bibinfo{journal}{\prd} \textbf{\bibinfo{volume}{62}}, \bibinfo{eid}{124022}
  (\bibinfo{year}{2000}), \eprint{gr-qc/0003032}.

\bibitem[{\citenamefont{{Buonanno} and {Damour}}(1999)}]{1999PhRvD..59h4006B}
\bibinfo{author}{\bibfnamefont{A.}~\bibnamefont{{Buonanno}}} \bibnamefont{and}
  \bibinfo{author}{\bibfnamefont{T.}~\bibnamefont{{Damour}}},
  \bibinfo{journal}{\prd} \textbf{\bibinfo{volume}{59}}, \bibinfo{eid}{084006}
  (\bibinfo{year}{1999}), \eprint{gr-qc/9811091}.

\bibitem[{\citenamefont{{Buonanno} and {Damour}}(2000)}]{2000PhRvD..62f4015B}
\bibinfo{author}{\bibfnamefont{A.}~\bibnamefont{{Buonanno}}} \bibnamefont{and}
  \bibinfo{author}{\bibfnamefont{T.}~\bibnamefont{{Damour}}},
  \bibinfo{journal}{\prd} \textbf{\bibinfo{volume}{62}}, \bibinfo{eid}{064015}
  (\bibinfo{year}{2000}), \eprint{gr-qc/0001013}.

\bibitem[{\citenamefont{{Liu} et~al.}(2018)\citenamefont{{Liu}, {Creswell},
  {von Hausegger}, {Jackson}, and {Naselsky}}}]{2018JCAP...02..013L}
\bibinfo{author}{\bibfnamefont{H.}~\bibnamefont{{Liu}}},
  \bibinfo{author}{\bibfnamefont{J.}~\bibnamefont{{Creswell}}},
  \bibinfo{author}{\bibfnamefont{S.}~\bibnamefont{{von Hausegger}}},
  \bibinfo{author}{\bibfnamefont{A.~D.} \bibnamefont{{Jackson}}},
  \bibnamefont{and}
  \bibinfo{author}{\bibfnamefont{P.}~\bibnamefont{{Naselsky}}},
  \bibinfo{journal}{\jcap} \textbf{\bibinfo{volume}{2}}, \bibinfo{eid}{013}
  (\bibinfo{year}{2018}), \eprint{1802.00340}.

\bibitem[{\citenamefont{{Naselsky} et~al.}(2016)\citenamefont{{Naselsky},
  {Jackson}, and {Liu}}}]{2016JCAP...08..029N}
\bibinfo{author}{\bibfnamefont{P.}~\bibnamefont{{Naselsky}}},
  \bibinfo{author}{\bibfnamefont{A.~D.} \bibnamefont{{Jackson}}},
  \bibnamefont{and} \bibinfo{author}{\bibfnamefont{H.}~\bibnamefont{{Liu}}},
  \bibinfo{journal}{\jcap} \textbf{\bibinfo{volume}{8}}, \bibinfo{eid}{029}
  (\bibinfo{year}{2016}), \eprint{1604.06211}.

\bibitem[{\citenamefont{{Liu} and {Jackson}}(2016)}]{2016JCAP...10..014L}
\bibinfo{author}{\bibfnamefont{H.}~\bibnamefont{{Liu}}} \bibnamefont{and}
  \bibinfo{author}{\bibfnamefont{A.~D.} \bibnamefont{{Jackson}}},
  \bibinfo{journal}{\jcap} \textbf{\bibinfo{volume}{10}}, \bibinfo{eid}{014}
  (\bibinfo{year}{2016}), \eprint{1609.08346}.

\bibitem[{\citenamefont{{Creswell} et~al.}(2017)\citenamefont{{Creswell}, {von
  Hausegger}, {Jackson}, {Liu}, and {Naselsky}}}]{2017JCAP...08..013C}
\bibinfo{author}{\bibfnamefont{J.}~\bibnamefont{{Creswell}}},
  \bibinfo{author}{\bibfnamefont{S.}~\bibnamefont{{von Hausegger}}},
  \bibinfo{author}{\bibfnamefont{A.~D.} \bibnamefont{{Jackson}}},
  \bibinfo{author}{\bibfnamefont{H.}~\bibnamefont{{Liu}}}, \bibnamefont{and}
  \bibinfo{author}{\bibfnamefont{P.}~\bibnamefont{{Naselsky}}},
  \bibinfo{journal}{\jcap} \textbf{\bibinfo{volume}{8}}, \bibinfo{eid}{013}
  (\bibinfo{year}{2017}), \eprint{1706.04191}.

\bibitem[{\citenamefont{{Ruffini} and {Wheeler}}(1971)}]{RuffiniWheeler71}
\bibinfo{author}{\bibfnamefont{R.}~\bibnamefont{{Ruffini}}} \bibnamefont{and}
  \bibinfo{author}{\bibfnamefont{J.~A.} \bibnamefont{{Wheeler}}}, in
  \emph{\bibinfo{booktitle}{Proceedings of the Cortona Symposium on Weak
  Interactions}}, edited by
  \bibinfo{editor}{\bibfnamefont{L.}~\bibnamefont{{Radicati}}}
  (\bibinfo{year}{1971}).

\bibitem[{\citenamefont{{Rees} et~al.}(1974)\citenamefont{{Rees}, {Ruffini},
  and {Wheeler}}}]{1974bhgw.book.....R}
\bibinfo{author}{\bibfnamefont{M.}~\bibnamefont{{Rees}}},
  \bibinfo{author}{\bibfnamefont{R.}~\bibnamefont{{Ruffini}}},
  \bibnamefont{and} \bibinfo{author}{\bibfnamefont{J.~A.}
  \bibnamefont{{Wheeler}}}, \emph{\bibinfo{title}{{Black holes, gravitational
  waves and cosmology}}} (\bibinfo{publisher}{New York: Gordon and Breach
  Science Publishers Inc.}, \bibinfo{year}{1974}).

\bibitem[{\citenamefont{{Zerilli}}(1970{\natexlab{a}})}]{1970PhRvL..24..737Z}
\bibinfo{author}{\bibfnamefont{F.~J.} \bibnamefont{{Zerilli}}},
  \bibinfo{journal}{Physical Review Letters} \textbf{\bibinfo{volume}{24}},
  \bibinfo{pages}{737} (\bibinfo{year}{1970}{\natexlab{a}}).

\bibitem[{\citenamefont{{Zerilli}}(1970{\natexlab{b}})}]{1970PhRvD...2.2141Z}
\bibinfo{author}{\bibfnamefont{F.~J.} \bibnamefont{{Zerilli}}},
  \bibinfo{journal}{\prd} \textbf{\bibinfo{volume}{2}}, \bibinfo{pages}{2141}
  (\bibinfo{year}{1970}{\natexlab{b}}).

\bibitem[{\citenamefont{{Regge} and {Wheeler}}(1957)}]{1957PhRv..108.1063R}
\bibinfo{author}{\bibfnamefont{T.}~\bibnamefont{{Regge}}} \bibnamefont{and}
  \bibinfo{author}{\bibfnamefont{J.~A.} \bibnamefont{{Wheeler}}},
  \bibinfo{journal}{Physical Review} \textbf{\bibinfo{volume}{108}},
  \bibinfo{pages}{1063} (\bibinfo{year}{1957}).

\bibitem[{\citenamefont{{Davis} et~al.}(1971)\citenamefont{{Davis}, {Ruffini},
  {Press}, and {Price}}}]{1971PhRvL..27.1466D}
\bibinfo{author}{\bibfnamefont{M.}~\bibnamefont{{Davis}}},
  \bibinfo{author}{\bibfnamefont{R.}~\bibnamefont{{Ruffini}}},
  \bibinfo{author}{\bibfnamefont{W.~H.} \bibnamefont{{Press}}},
  \bibnamefont{and} \bibinfo{author}{\bibfnamefont{R.~H.}
  \bibnamefont{{Price}}}, \bibinfo{journal}{Physical Review Letters}
  \textbf{\bibinfo{volume}{27}}, \bibinfo{pages}{1466} (\bibinfo{year}{1971}).

\bibitem[{\citenamefont{{Davis}
  et~al.}(1972{\natexlab{a}})\citenamefont{{Davis}, {Ruffini}, and
  {Tiomno}}}]{1972PhRvD...5.2932D}
\bibinfo{author}{\bibfnamefont{M.}~\bibnamefont{{Davis}}},
  \bibinfo{author}{\bibfnamefont{R.}~\bibnamefont{{Ruffini}}},
  \bibnamefont{and} \bibinfo{author}{\bibfnamefont{J.}~\bibnamefont{{Tiomno}}},
  \bibinfo{journal}{\prd} \textbf{\bibinfo{volume}{5}}, \bibinfo{pages}{2932}
  (\bibinfo{year}{1972}{\natexlab{a}}).

\bibitem[{\citenamefont{{Ruffini}}(1973{\natexlab{a}})}]{1973PhRvD...7..972R}
\bibinfo{author}{\bibfnamefont{R.}~\bibnamefont{{Ruffini}}},
  \bibinfo{journal}{\prd} \textbf{\bibinfo{volume}{7}}, \bibinfo{pages}{972}
  (\bibinfo{year}{1973}{\natexlab{a}}).

\bibitem[{\citenamefont{{Ruffini} and {Sasaki}}(1981)}]{1981PThPh..66.1627R}
\bibinfo{author}{\bibfnamefont{R.}~\bibnamefont{{Ruffini}}} \bibnamefont{and}
  \bibinfo{author}{\bibfnamefont{M.}~\bibnamefont{{Sasaki}}},
  \bibinfo{journal}{Progress of Theoretical Physics}
  \textbf{\bibinfo{volume}{66}}, \bibinfo{pages}{1627} (\bibinfo{year}{1981}).

\bibitem[{\citenamefont{{Teukolsky}}(1973)}]{1973ApJ...185..635T}
\bibinfo{author}{\bibfnamefont{S.~A.} \bibnamefont{{Teukolsky}}},
  \bibinfo{journal}{\apj} \textbf{\bibinfo{volume}{185}}, \bibinfo{pages}{635}
  (\bibinfo{year}{1973}).

\bibitem[{\citenamefont{{Detweiler} and
  {Szedenits}}(1979)}]{1979ApJ...231..211D}
\bibinfo{author}{\bibfnamefont{S.~L.} \bibnamefont{{Detweiler}}}
  \bibnamefont{and}
  \bibinfo{author}{\bibfnamefont{E.}~\bibnamefont{{Szedenits}},
  \bibfnamefont{Jr.}}, \bibinfo{journal}{\apj} \textbf{\bibinfo{volume}{231}},
  \bibinfo{pages}{211} (\bibinfo{year}{1979}).

\bibitem[{\citenamefont{{Chandrasekhar} and
  {Detweiler}}(1976)}]{1976RSPSA.350..165C}
\bibinfo{author}{\bibfnamefont{S.}~\bibnamefont{{Chandrasekhar}}}
  \bibnamefont{and}
  \bibinfo{author}{\bibfnamefont{S.}~\bibnamefont{{Detweiler}}},
  \bibinfo{journal}{Proceedings of the Royal Society of London Series A}
  \textbf{\bibinfo{volume}{350}}, \bibinfo{pages}{165} (\bibinfo{year}{1976}).

\bibitem[{\citenamefont{{Sasaki} and
  {Nakamura}}(1982{\natexlab{a}})}]{1982PhLA...89...68S}
\bibinfo{author}{\bibfnamefont{M.}~\bibnamefont{{Sasaki}}} \bibnamefont{and}
  \bibinfo{author}{\bibfnamefont{T.}~\bibnamefont{{Nakamura}}},
  \bibinfo{journal}{Physics Letters A} \textbf{\bibinfo{volume}{89}},
  \bibinfo{pages}{68} (\bibinfo{year}{1982}{\natexlab{a}}).

\bibitem[{\citenamefont{{Sasaki} and
  {Nakamura}}(1982{\natexlab{b}})}]{1982PThPh..67.1788S}
\bibinfo{author}{\bibfnamefont{M.}~\bibnamefont{{Sasaki}}} \bibnamefont{and}
  \bibinfo{author}{\bibfnamefont{T.}~\bibnamefont{{Nakamura}}},
  \bibinfo{journal}{Progress of Theoretical Physics}
  \textbf{\bibinfo{volume}{67}}, \bibinfo{pages}{1788}
  (\bibinfo{year}{1982}{\natexlab{b}}).

\bibitem[{\citenamefont{{Kojima} and {Nakamura}}(1984)}]{1984PThPh..71...79K}
\bibinfo{author}{\bibfnamefont{Y.}~\bibnamefont{{Kojima}}} \bibnamefont{and}
  \bibinfo{author}{\bibfnamefont{T.}~\bibnamefont{{Nakamura}}},
  \bibinfo{journal}{Progress of Theoretical Physics}
  \textbf{\bibinfo{volume}{71}}, \bibinfo{pages}{79} (\bibinfo{year}{1984}).

\bibitem[{\citenamefont{{Weber}}(1969)}]{1969PhRvL..22.1320W}
\bibinfo{author}{\bibfnamefont{J.}~\bibnamefont{{Weber}}},
  \bibinfo{journal}{Physical Review Letters} \textbf{\bibinfo{volume}{22}},
  \bibinfo{pages}{1320} (\bibinfo{year}{1969}).

\bibitem[{\citenamefont{{Misner}}(1972)}]{1972PhRvL..28..994M}
\bibinfo{author}{\bibfnamefont{C.~W.} \bibnamefont{{Misner}}},
  \bibinfo{journal}{Physical Review Letters} \textbf{\bibinfo{volume}{28}},
  \bibinfo{pages}{994} (\bibinfo{year}{1972}).

\bibitem[{\citenamefont{{Misner} et~al.}(1972)\citenamefont{{Misner}, {Breuer},
  {Brill}, {Chrzanowski}, {Hughes}, and {Pereira}}}]{1972PhRvL..28..998M}
\bibinfo{author}{\bibfnamefont{C.~W.} \bibnamefont{{Misner}}},
  \bibinfo{author}{\bibfnamefont{R.~A.} \bibnamefont{{Breuer}}},
  \bibinfo{author}{\bibfnamefont{D.~R.} \bibnamefont{{Brill}}},
  \bibinfo{author}{\bibfnamefont{P.~L.} \bibnamefont{{Chrzanowski}}},
  \bibinfo{author}{\bibfnamefont{H.~G.} \bibnamefont{{Hughes}}},
  \bibnamefont{and} \bibinfo{author}{\bibfnamefont{C.~M.}
  \bibnamefont{{Pereira}}}, \bibinfo{journal}{Physical Review Letters}
  \textbf{\bibinfo{volume}{28}}, \bibinfo{pages}{998} (\bibinfo{year}{1972}).

\bibitem[{\citenamefont{{Davis}
  et~al.}(1972{\natexlab{b}})\citenamefont{{Davis}, {Ruffini}, {Tiomno}, and
  {Zerilli}}}]{1972PhRvL..28.1352D}
\bibinfo{author}{\bibfnamefont{M.}~\bibnamefont{{Davis}}},
  \bibinfo{author}{\bibfnamefont{R.}~\bibnamefont{{Ruffini}}},
  \bibinfo{author}{\bibfnamefont{J.}~\bibnamefont{{Tiomno}}}, \bibnamefont{and}
  \bibinfo{author}{\bibfnamefont{F.}~\bibnamefont{{Zerilli}}},
  \bibinfo{journal}{Physical Review Letters} \textbf{\bibinfo{volume}{28}},
  \bibinfo{pages}{1352} (\bibinfo{year}{1972}{\natexlab{b}}).

\bibitem[{\citenamefont{{Cutler} et~al.}(1993)\citenamefont{{Cutler}, {Finn},
  {Poisson}, and {Sussman}}}]{1993PhRvD..47.1511C}
\bibinfo{author}{\bibfnamefont{C.}~\bibnamefont{{Cutler}}},
  \bibinfo{author}{\bibfnamefont{L.~S.} \bibnamefont{{Finn}}},
  \bibinfo{author}{\bibfnamefont{E.}~\bibnamefont{{Poisson}}},
  \bibnamefont{and} \bibinfo{author}{\bibfnamefont{G.~J.}
  \bibnamefont{{Sussman}}}, \bibinfo{journal}{\prd}
  \textbf{\bibinfo{volume}{47}}, \bibinfo{pages}{1511} (\bibinfo{year}{1993}).

\bibitem[{\citenamefont{{Damour} et~al.}(2009)\citenamefont{{Damour}, {Iyer},
  and {Nagar}}}]{2009PhRvD..79f4004D}
\bibinfo{author}{\bibfnamefont{T.}~\bibnamefont{{Damour}}},
  \bibinfo{author}{\bibfnamefont{B.~R.} \bibnamefont{{Iyer}}},
  \bibnamefont{and} \bibinfo{author}{\bibfnamefont{A.}~\bibnamefont{{Nagar}}},
  \bibinfo{journal}{\prd} \textbf{\bibinfo{volume}{79}}, \bibinfo{eid}{064004}
  (\bibinfo{year}{2009}), \eprint{0811.2069}.

\bibitem[{\citenamefont{{Tanaka} et~al.}(1993)\citenamefont{{Tanaka},
  {Shibata}, {Sasaki}, {Tagoshi}, and {Nakamura}}}]{1993PThPh..90...65T}
\bibinfo{author}{\bibfnamefont{T.}~\bibnamefont{{Tanaka}}},
  \bibinfo{author}{\bibfnamefont{M.}~\bibnamefont{{Shibata}}},
  \bibinfo{author}{\bibfnamefont{M.}~\bibnamefont{{Sasaki}}},
  \bibinfo{author}{\bibfnamefont{H.}~\bibnamefont{{Tagoshi}}},
  \bibnamefont{and}
  \bibinfo{author}{\bibfnamefont{T.}~\bibnamefont{{Nakamura}}},
  \bibinfo{journal}{Progress of Theoretical Physics}
  \textbf{\bibinfo{volume}{90}}, \bibinfo{pages}{65} (\bibinfo{year}{1993}).

\bibitem[{\citenamefont{{Detweiler}}(1978)}]{1978ApJ...225..687D}
\bibinfo{author}{\bibfnamefont{S.~L.} \bibnamefont{{Detweiler}}},
  \bibinfo{journal}{\apj} \textbf{\bibinfo{volume}{225}}, \bibinfo{pages}{687}
  (\bibinfo{year}{1978}).

\bibitem[{\citenamefont{{Shibata}}(1993{\natexlab{a}})}]{1993PhRvD..48..663S}
\bibinfo{author}{\bibfnamefont{M.}~\bibnamefont{{Shibata}}},
  \bibinfo{journal}{\prd} \textbf{\bibinfo{volume}{48}}, \bibinfo{pages}{663}
  (\bibinfo{year}{1993}{\natexlab{a}}).

\bibitem[{\citenamefont{{Shibata}}(1993{\natexlab{b}})}]{1993PThPh..90..595S}
\bibinfo{author}{\bibfnamefont{M.}~\bibnamefont{{Shibata}}},
  \bibinfo{journal}{Progress of Theoretical Physics}
  \textbf{\bibinfo{volume}{90}}, \bibinfo{pages}{595}
  (\bibinfo{year}{1993}{\natexlab{b}}).

\bibitem[{\citenamefont{{Bardeen} et~al.}(1972)\citenamefont{{Bardeen},
  {Press}, and {Teukolsky}}}]{1972ApJ...178..347B}
\bibinfo{author}{\bibfnamefont{J.~M.} \bibnamefont{{Bardeen}}},
  \bibinfo{author}{\bibfnamefont{W.~H.} \bibnamefont{{Press}}},
  \bibnamefont{and} \bibinfo{author}{\bibfnamefont{S.~A.}
  \bibnamefont{{Teukolsky}}}, \bibinfo{journal}{\apj}
  \textbf{\bibinfo{volume}{178}}, \bibinfo{pages}{347} (\bibinfo{year}{1972}).

\bibitem[{\citenamefont{{Teukolsky}}(1972)}]{1972PhRvL..29.1114T}
\bibinfo{author}{\bibfnamefont{S.~A.} \bibnamefont{{Teukolsky}}},
  \bibinfo{journal}{Physical Review Letters} \textbf{\bibinfo{volume}{29}},
  \bibinfo{pages}{1114} (\bibinfo{year}{1972}).

\bibitem[{\citenamefont{{Teukolsky} and {Press}}(1974)}]{1974ApJ...193..443T}
\bibinfo{author}{\bibfnamefont{S.~A.} \bibnamefont{{Teukolsky}}}
  \bibnamefont{and} \bibinfo{author}{\bibfnamefont{W.~H.}
  \bibnamefont{{Press}}}, \bibinfo{journal}{\apj}
  \textbf{\bibinfo{volume}{193}}, \bibinfo{pages}{443} (\bibinfo{year}{1974}).

\bibitem[{\citenamefont{{Hughes}}(2000)}]{2000PhRvD..61h4004H}
\bibinfo{author}{\bibfnamefont{S.~A.} \bibnamefont{{Hughes}}},
  \bibinfo{journal}{\prd} \textbf{\bibinfo{volume}{61}}, \bibinfo{eid}{084004}
  (\bibinfo{year}{2000}), \eprint{gr-qc/9910091}.

\bibitem[{\citenamefont{Nakano et~al.}(2016)\citenamefont{Nakano, Sago, Tanaka,
  and Nakamura}}]{nakano2016}
\bibinfo{author}{\bibfnamefont{H.}~\bibnamefont{Nakano}},
  \bibinfo{author}{\bibfnamefont{N.}~\bibnamefont{Sago}},
  \bibinfo{author}{\bibfnamefont{T.}~\bibnamefont{Tanaka}}, \bibnamefont{and}
  \bibinfo{author}{\bibfnamefont{T.}~\bibnamefont{Nakamura}},
  \bibinfo{journal}{Progress of Theoretical and Experimental Physics}
  \textbf{\bibinfo{volume}{2016}}, \bibinfo{pages}{083E01}
  (\bibinfo{year}{2016}).

\bibitem[{\citenamefont{{Gralla} et~al.}(2016)\citenamefont{{Gralla}, {Hughes},
  and {Warburton}}}]{2016CQGra..33o5002G}
\bibinfo{author}{\bibfnamefont{S.~E.} \bibnamefont{{Gralla}}},
  \bibinfo{author}{\bibfnamefont{S.~A.} \bibnamefont{{Hughes}}},
  \bibnamefont{and}
  \bibinfo{author}{\bibfnamefont{N.}~\bibnamefont{{Warburton}}},
  \bibinfo{journal}{Classical and Quantum Gravity}
  \textbf{\bibinfo{volume}{33}}, \bibinfo{eid}{155002} (\bibinfo{year}{2016}),
  \eprint{1603.01221}.

\bibitem[{\citenamefont{{Jantzen} et~al.}(1992)\citenamefont{{Jantzen},
  {Carini}, and {Bini}}}]{1992AnPhy.215....1J}
\bibinfo{author}{\bibfnamefont{R.~T.} \bibnamefont{{Jantzen}}},
  \bibinfo{author}{\bibfnamefont{P.}~\bibnamefont{{Carini}}}, \bibnamefont{and}
  \bibinfo{author}{\bibfnamefont{D.}~\bibnamefont{{Bini}}},
  \bibinfo{journal}{Annals of Physics} \textbf{\bibinfo{volume}{215}},
  \bibinfo{pages}{1} (\bibinfo{year}{1992}), \eprint{gr-qc/0106043}.

\bibitem[{\citenamefont{{Poisson} et~al.}(2011)\citenamefont{{Poisson},
  {Pound}, and {Vega}}}]{2011LRR....14....7P}
\bibinfo{author}{\bibfnamefont{E.}~\bibnamefont{{Poisson}}},
  \bibinfo{author}{\bibfnamefont{A.}~\bibnamefont{{Pound}}}, \bibnamefont{and}
  \bibinfo{author}{\bibfnamefont{I.}~\bibnamefont{{Vega}}},
  \bibinfo{journal}{Living Reviews in Relativity}
  \textbf{\bibinfo{volume}{14}}, \bibinfo{eid}{7} (\bibinfo{year}{2011}),
  \eprint{1102.0529}.

\bibitem[{\citenamefont{{Fitchett} and
  {Detweiler}}(1984)}]{1984MNRAS.211..933F}
\bibinfo{author}{\bibfnamefont{M.~J.} \bibnamefont{{Fitchett}}}
  \bibnamefont{and}
  \bibinfo{author}{\bibfnamefont{S.}~\bibnamefont{{Detweiler}}},
  \bibinfo{journal}{\mnras} \textbf{\bibinfo{volume}{211}},
  \bibinfo{pages}{933} (\bibinfo{year}{1984}).

\bibitem[{\citenamefont{{Finn} and {Thorne}}(2000)}]{2000PhRvD..62l4021F}
\bibinfo{author}{\bibfnamefont{L.~S.} \bibnamefont{{Finn}}} \bibnamefont{and}
  \bibinfo{author}{\bibfnamefont{K.~S.} \bibnamefont{{Thorne}}},
  \bibinfo{journal}{\prd} \textbf{\bibinfo{volume}{62}}, \bibinfo{eid}{124021}
  (\bibinfo{year}{2000}), \eprint{gr-qc/0007074}.

\bibitem[{\citenamefont{{Hughes}}(2001)}]{2001PhRvD..64f4004H}
\bibinfo{author}{\bibfnamefont{S.~A.} \bibnamefont{{Hughes}}},
  \bibinfo{journal}{\prd} \textbf{\bibinfo{volume}{64}}, \bibinfo{eid}{064004}
  (\bibinfo{year}{2001}), \eprint{gr-qc/0104041}.

\bibitem[{\citenamefont{{Damour}}(2001)}]{2001PhRvD..64l4013D}
\bibinfo{author}{\bibfnamefont{T.}~\bibnamefont{{Damour}}},
  \bibinfo{journal}{\prd} \textbf{\bibinfo{volume}{64}},
  \bibinfo{pages}{124013} (\bibinfo{year}{2001}), \eprint{gr-qc/0103018}.

\bibitem[{\citenamefont{{Han} and {Cao}}(2011)}]{2011PhRvD..84d4014H}
\bibinfo{author}{\bibfnamefont{W.-B.} \bibnamefont{{Han}}} \bibnamefont{and}
  \bibinfo{author}{\bibfnamefont{Z.}~\bibnamefont{{Cao}}},
  \bibinfo{journal}{\prd} \textbf{\bibinfo{volume}{84}}, \bibinfo{eid}{044014}
  (\bibinfo{year}{2011}), \eprint{1108.0995}.

\bibitem[{\citenamefont{{Pound} and {Poisson}}(2008)}]{2008PhRvD..77d4013P}
\bibinfo{author}{\bibfnamefont{A.}~\bibnamefont{{Pound}}} \bibnamefont{and}
  \bibinfo{author}{\bibfnamefont{E.}~\bibnamefont{{Poisson}}},
  \bibinfo{journal}{\prd} \textbf{\bibinfo{volume}{77}}, \bibinfo{eid}{044013}
  (\bibinfo{year}{2008}), \eprint{0708.3033}.

\bibitem[{\citenamefont{{Gair} et~al.}(2011)\citenamefont{{Gair}, {Flanagan},
  {Drasco}, {Hinderer}, and {Babak}}}]{2011PhRvD..83d4037G}
\bibinfo{author}{\bibfnamefont{J.~R.} \bibnamefont{{Gair}}},
  \bibinfo{author}{\bibfnamefont{{\'E}.~{\'E}.} \bibnamefont{{Flanagan}}},
  \bibinfo{author}{\bibfnamefont{S.}~\bibnamefont{{Drasco}}},
  \bibinfo{author}{\bibfnamefont{T.}~\bibnamefont{{Hinderer}}},
  \bibnamefont{and} \bibinfo{author}{\bibfnamefont{S.}~\bibnamefont{{Babak}}},
  \bibinfo{journal}{\prd} \textbf{\bibinfo{volume}{83}}, \bibinfo{eid}{044037}
  (\bibinfo{year}{2011}), \eprint{1012.5111}.

\bibitem[{\citenamefont{{Taracchini} et~al.}(2014)\citenamefont{{Taracchini},
  {Buonanno}, {Khanna}, and {Hughes}}}]{2014PhRvD..90h4025T}
\bibinfo{author}{\bibfnamefont{A.}~\bibnamefont{{Taracchini}}},
  \bibinfo{author}{\bibfnamefont{A.}~\bibnamefont{{Buonanno}}},
  \bibinfo{author}{\bibfnamefont{G.}~\bibnamefont{{Khanna}}}, \bibnamefont{and}
  \bibinfo{author}{\bibfnamefont{S.~A.} \bibnamefont{{Hughes}}},
  \bibinfo{journal}{\prd} \textbf{\bibinfo{volume}{90}}, \bibinfo{eid}{084025}
  (\bibinfo{year}{2014}), \eprint{1404.1819}.

\bibitem[{\citenamefont{{Cutler} et~al.}(1994)\citenamefont{{Cutler},
  {Kennefick}, and {Poisson}}}]{1994PhRvD..50.3816C}
\bibinfo{author}{\bibfnamefont{C.}~\bibnamefont{{Cutler}}},
  \bibinfo{author}{\bibfnamefont{D.}~\bibnamefont{{Kennefick}}},
  \bibnamefont{and}
  \bibinfo{author}{\bibfnamefont{E.}~\bibnamefont{{Poisson}}},
  \bibinfo{journal}{\prd} \textbf{\bibinfo{volume}{50}}, \bibinfo{pages}{3816}
  (\bibinfo{year}{1994}).

\bibitem[{\citenamefont{{Ruffini}}(1973{\natexlab{b}})}]{1973blho.conf..451R}
\bibinfo{author}{\bibfnamefont{R.}~\bibnamefont{{Ruffini}}}, in
  \emph{\bibinfo{booktitle}{Black Holes (Les Astres Occlus)}}, edited by
  \bibinfo{editor}{\bibfnamefont{C.}~\bibnamefont{{Dewitt}}} \bibnamefont{and}
  \bibinfo{editor}{\bibfnamefont{B.~S.} \bibnamefont{{Dewitt}}}
  (\bibinfo{year}{1973}{\natexlab{b}}), pp. \bibinfo{pages}{451--546}.

\end{thebibliography}

\end{document}